\documentclass[useAMS,usenatbib]{mn2e} 

\usepackage{times,graphicx,amssymb}
\usepackage[draft]{hyperref}

\newcommand{\ion}[2]{{#1~\small#2}}

\newcommand{\HI}{$\rm H\,{\sevensize I}$}

\newcommand{\lya}{Ly$\alpha$}
\newcommand{\ha}{H$\alpha$}

\newcommand{\sbcgs}{$\rm erg~s^{-1}~cm^{-2}~\AA^{-1}~arcsec^{-2}$}
\newcommand{\sbline}{$\rm erg~s^{-1}~cm^{-2}~arcsec^{-2}$}

\newcommand{\gal}{UGC~7321}
\newcommand{\cubex}{{\sc CubEx}}
\newcommand{\zapline}{{\sc zap}}

\title[A measurement of the $z=0$ UVB.]{A measurement of the $z=0$ UV background from H$\alpha$ fluorescence.}

\author[Fumagalli et al.]{Michele Fumagalli$^{1,2}$\thanks{E-mail: michele.fumagalli@durham.ac.uk},
  Francesco Haardt$^{3,4}$, Tom Theuns$^{1}$, Simon L. Morris$^{2}$,\and Sebastiano Cantalupo$^{5}$,
  Piero Madau$^{6}$, Matteo Fossati$^{7,8}$\\
  $^{1}$Institute for Computational Cosmology,Durham University, South Road, Durham, DH1 3LE, UK \\
  $^{2}$Centre for Extragalactic Astronomy, Durham University, South Road, Durham, DH1 3LE, UK \\
  $^{3}$DiSAT, Universit\`a degli Studi dell'Insubria, Via Valleggio 11, 22100 Como, Italy\\
  $^{4}$INFN, Sezione di Milano-Bicocca, Piazza della Scienza 3, 20126 Milano, Italy\\
  $^{5}$Institute for Astronomy, ETH Zurich, Wolfgang-Pauli-Strasse 27, 8093 Zurich, Switzerland\\
  $^{6}$Department of Astronomy \& Astrophysics, University of California, 1156 High Street, Santa Cruz, CA 95064\\
  $^{7}$Universit{\"a}ts-Sternwarte M{\"u}nchen, Scheinerstrasse 1, 81679 M{\"unchen}, Germany \\
  $^{8}$Max-Planck-Institut f{\"u}r Extraterrestrische Physik, Giessenbachstrasse, 85748 Garching, Germany\\
}

\begin{document}


\pagerange{\pageref{firstpage}--\pageref{lastpage}} \pubyear{\the\year}

\maketitle

\label{firstpage}

\begin{abstract}
  We report the detection of extended H$\alpha$ emission from the tip of the \HI\ disk of the nearby edge-on galaxy UGC~7321, observed with the Multi Unit Spectroscopic Explorer (MUSE) instrument at the Very Large Telescope. The H$\alpha$ surface brightness fades rapidly where the \HI\ column density drops below $N_{\rm HI}\sim 10^{19}$~cm$^{-2}$, consistent with fluorescence arising at the ionisation front from gas that is photoionized by the extragalactic ultraviolet background (UVB). The  surface brightness measured at this location is  $(1.2\pm 0.5)\times 10^{-19}~$\sbline, where the error is mostly systematic and results from the proximity of the signal to the edge of the MUSE field of view, and from the presence of a  sky line next to the redshifted H$\alpha$ wavelength. By combining the H$\alpha$ and the \HI\ 21~cm maps with a radiative transfer calculation of an exponential disk illuminated by the UVB, we derive a value for the \HI\ photoionization rate of $\Gamma_{\rm HI} \sim (6-8)\times 10^{-14}~\rm s^{-1}$. This value is consistent with transmission statistics of the Ly$\alpha$ forest and with recent models of a UVB which is dominated by quasars. 
\end{abstract}


\section{Introduction}

Massive stars and active galactic nuclei (AGN) in galaxies produce copious amounts of ultraviolet (UV) radiation. A fraction of these UV photons escape from the interstellar medium (ISM) of the host galaxy into the intergalactic medium (IGM), building up an extragalactic UV background (UVB). Following reionization, this UVB keeps the bulk of the IGM ionised \citep[e.g.][]{Gunn65,bol07}, regulates its temperature \citep[e.g.][]{the02}, and sets a characteristic virial temperature below which halos do not form galaxies \citep[e.g.][]{oka08}. The UVB is therefore an important ingredient in models of galaxy formation. Moreover, the UVB encodes the cumulative history of star formation and AGN activity, and depends on the redshift and luminosity-dependent escape fractions of galaxies \citep[e.g.][]{haa96}. A detailed understanding of the time evolution of the spectral shape and intensity of the UVB (hereafter $J_{\nu}(z)$) is of critical importance in many areas of astrophysics.

The amplitude of the UVB at redshifts $z\sim 2-3$ is expected to be more than ten times the present-day value \citep{haa12}, and three methods have been used to attempt to measure $J_{\nu}$ at these redshifts. Firstly, a background of \HI\ ionising photons will result in recombination radiation, such as Ly$\alpha$, when such photons impinge on optically-thick \HI\ clouds \citep[e.g.][]{gou96,can05}. However, searches for this Ly$\alpha$ \lq fluorescence\rq~ have remained inconclusive \citep[e.g.][]{bun98,rau08}. The expected intrinsic surface brightness (SB) is low, and the signal is furthermore significantly lowered by cosmological redshifting, making this measurement very challenging. Secondly, $J_{\nu}$ can be constrained by determining out to which distance a luminous source, such as a quasar,
	outshines the UVB, via the so-called \lq proximity effect\rq\  \citep[e.g.][]{mur86,baj88,Rollinde05}. However,
	the value inferred for $J_{\nu}$ depends on other properties of the system that are difficult to constrain, such as the time-dependence
	of the luminosity of the source, and the temperature and density structure of its surrounding medium \citep{fau08b,pro13}. Third, constraints on $J_{\nu}$ can be derived by comparing the observed transmission statistics of the Ly$\alpha$ forest to those
	measured in hydrodynamic simulations \cite[e.g.][]{Rauch97}. This method, which currently offers the primary constraints on
        $J_{\nu}$, suffers from systematic uncertainties, because
	the transmission statistics also depend on the relatively poorly-known temperature-density relation of the photoionized IGM
	\citep[e.g.][]{fau08,bec13,bol05}.

In the low-redshift Universe, at $z\lesssim 1$, Ly$\alpha$ transmission statistics also provide the best current constraints on $J_{\nu}$ \citep[e.g.][]{kol14,shu15,kha15,vie16}, but observing the Ly$\alpha$ forest requires UV-spectroscopy from space. The detection of Ly$\alpha$ fluorescence is challenging because the amplitude of the UVB is low and the IGM is more diffuse at these redshifts when compared to $z\sim 2-3$.
Interestingly, fluorescence could also be detected in H$\alpha$, by observing the ionisation front of neutral \HI\ clouds photoionized by the UVB \citep[][]{vog95,don95,wey01}, or in the outskirts  of the \HI\ disks of galaxies \citep[e.g.][]{mal93,dov94,bla97,cir99,mad01}.
Using this technique, \citet{ada11} targeted the nearby edge-on galaxy \gal, obtaining
an upper limit on $\Gamma_{\rm HI}$, which is the \HI\ photoionization rate of the UVB\footnote{The photoionization rate is $\Gamma_{\rm HI}=\int^{\infty}_{\nu_0} (4\pi J_{\nu}/h\nu)\,\sigma_{\rm HI}(\nu)\,d\nu$, where $\sigma_{\rm HI}$ is the photoionization cross section and $\nu_0$ is the
frequency corresponding to the ionisation potential of hydrogen.}.
The same group also reported a detection\footnote{The announcement of this result is also available at
  \url{http://iactalks.iac.es/talks/view/393}}, which has not been published at the time of writing \citep[see][]{uso12}.

In the absence of firm observational constraints on $J_{\nu}$, the current parametrisation of the UVB relies mostly on radiative transfer calculations that follow the build-up of the UVB accounting for sources and sinks of radiation. These models have input parameters that are difficult to measure, such as the emissivity and escape fraction of ionising photons from massive stars and AGN in galaxies, and the distribution of \HI\ absorbers \citep{haa96,shu99,fau09,haa12}. Therefore, different models predict values of $\Gamma_{\rm HI}$ that differ by factors of a few, primarily because the observational data that enter the modelling are not well known. Such a relatively large uncertainty in $\Gamma_{\rm HI}$
	then impacts the reliability of other predictions, for example the outcome of cosmological hydrodynamic simulations
	\citep[see e.g.][]{kol14,oor16}. For example, \cite{kol14} argue that the  value of $J_{\nu}$ predicted by
	\citet{haa12} underestimates the UVB at $z\sim 0$ by a factor up to five compared to what is required by the transmission
	statistics of the low-redshift \lya\ forest.  Follow-up work confirms this discrepancy, although revising downward its severity
	\citep[e.g.][]{shu15,kha15,vie16}.

In this paper we describe the results from new observations designed to measure $J_{\rm \nu}$ through the experiment proposed by \citet{ada11}, who attempted to measure the UVB intensity by searching for the \ha\ recombination line arising from gas that is photoionized by the UVB at the edge of the \HI\ disk in the nearby spiral galaxy \gal. The distance to this galaxy is $\sim 10~\rm Mpc$ and it has a mostly-unperturbed \HI\ disk seen edge-on, thus providing the ideal conditions for measuring the H$\alpha$ fluorescence induced by the UVB. A critical breakthrough enabling this experiment is the deployment of the Multi Unit Spectroscopic Explorer \citep[MUSE;][]{bac10} at the Very Large Telescope (VLT), which offers a  powerful combination of a relatively large field of view (FOV; $1\times1$ arcmin$^2$) and high throughput ($\sim 35\%$ at $\lambda\sim 6600~$\AA). Indeed, the capability to combine the large collecting area of VLT with an integral field spectrograph allows observers to create composites of $\gtrsim 10,000$ independent spectra, thus increasing by a factor $\gtrsim 100$ the sensitivity achievable with traditional long-slit spectrographs \citep[e.g.][]{rau08}. 
 
Here, we present results from a pilot MUSE observation, reporting a detection of extended \ha\ emission at the location 
	of the ionisation front inferred from photoionization models for \gal. The layout of this paper is as follows. In Sect. \ref{sec:obs} we describe the new observations and the reduction techniques, followed by the analysis of the data in Sect. \ref{sec:data}. In Sect. \ref{sec:uvb} we present updated photoionization modelling of \gal, through which we constrain the intensity of the UVB at $z\sim 0$. We summarise our results in Sect.~\ref{sec:end}, concluding with a discussion of how future observations can refine the measurement of the UVB intensity.

\section{Observations and data reduction}\label{sec:obs}

\begin{figure*}
  \centering
  \includegraphics[scale=0.65]{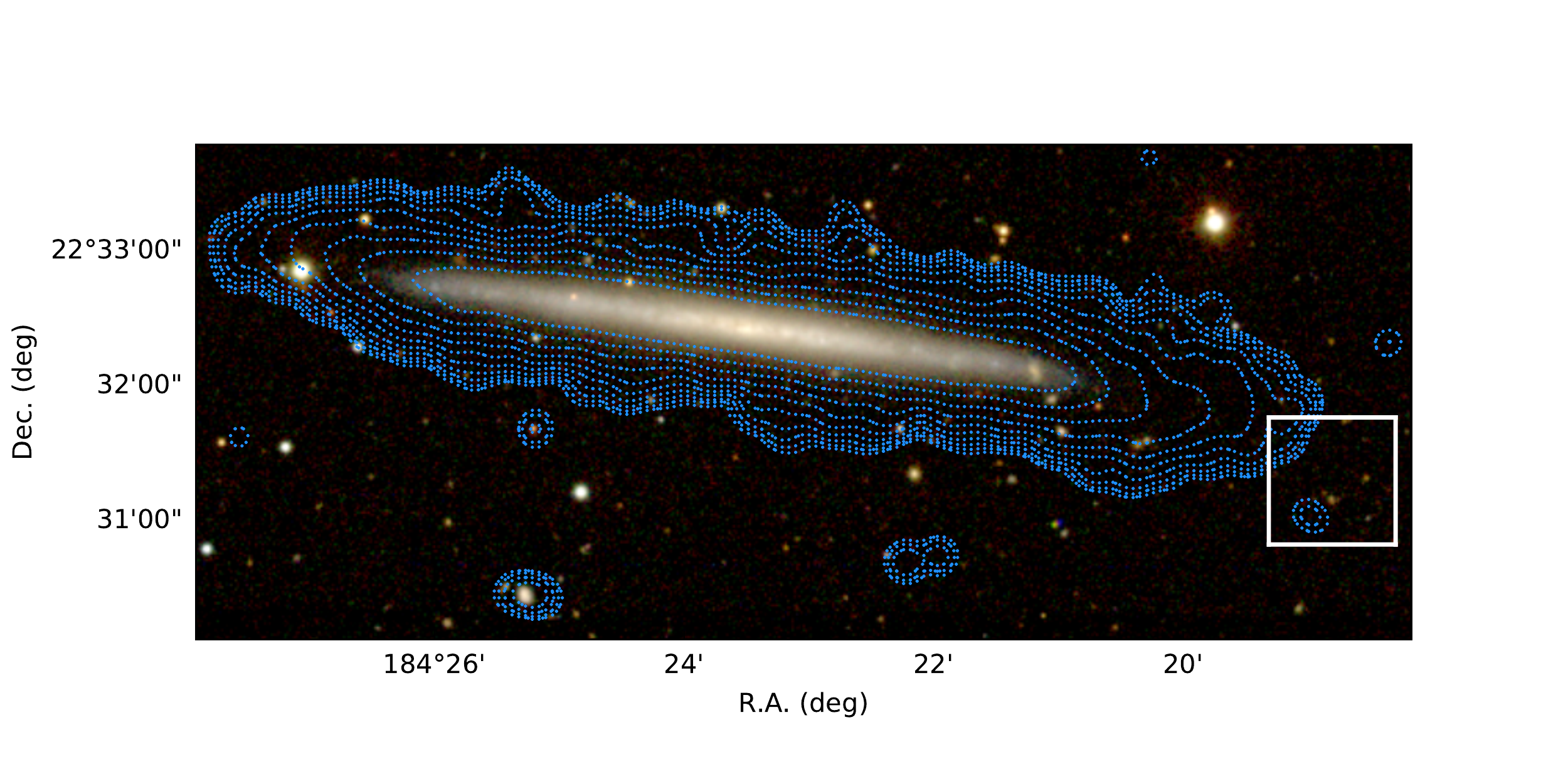}
  \caption{False-colour RGB image of \gal\ from SDSS imaging ($gri$), with \HI\ column density
    contours from \citet{ada11} in steps of $(1.0, 1.9, 3.6, 6.7, 12.6, 23.8, 44.8, 84.5)\times 10^{19}~\rm cm^{-2}$ for
    the outermost eight contours. The position of the MUSE FOV is shown in white.}\label{fig:galfov}
\end{figure*}

MUSE  observations  of \gal\  have been acquired as part of the programme ID 095.A-0090 (PI Fumagalli) between June 2015 and
January 2016 at the UT4 VLT. All observations have been completed in dark time, under clear or photometric conditions, with seeing 
$\lesssim 1.5~\rm arcsec$ and airmass $<1.6$. For these observations, we used the MUSE Wide Field Mode 
with slow guiding. A total of 14 exposures of 1465 s each, totalling 5.7 h on target, were acquired
at the position $\alpha{\rm (J2000)}=~$12:17:15.3 and $\delta{\rm (J2000)}=+$22:31:16.9 with small offsets and 90 degree
rotations in between exposures. Figure \ref{fig:galfov} shows an RGB image of \gal\
with the position of the MUSE FOV and \HI\ contours from \citet{ada11} \citep[see also][]{uso03}.
The location of the pointing was chosen to overlap with the region
where the SB was expected to be maximal according to the model of \citet{ada11},
while sampling a mostly blank sky region in the bottom part of the field of view as well (see Figure \ref{fig:musefov}).

Individual exposures have been reduced using the ESO MUSE pipeline \citep[v1.6.2;][]{wei14} which applies standard 
calibrations to the raw data, including bias subtraction, flat fielding, flux and wavelength calibrations,
and baryocentric corrections.

After the individual exposures have been processed with basic reduction techniques, we produce three final data sets for the
subsequent analysis using: i) the ESO pipeline; ii) the {\sc CubExtractor} package (\cubex, Cantalupo in prep.)
following the procedures described in \citet{bor16} and \citet{fum16muse}; iii) a custom Python post-processing
pipeline and the Zurich Atmosphere Purge (\zapline) package \citep{sot16}.
Each of these three methods applies independent algorithms for the sky subtraction
and, in some cases, for additional illumination corrections, allowing us to further test the robustness of
our results with respect to different reduction techniques. 

In the following, fluxes recorded in the data cubes are converted into SB units
using the pixel size of $0.2\times 0.2~\rm arcsec^2$. We also apply a correction for Galactic extinction
in the direction of \gal, which we estimate to be $f_{\rm dust}=1.06 \pm 0.01$ from the Milky Way dust map \citep{sch11}. 
As described in \citet{ada11}, the internal extinction of \gal\ is believed to have negligible effects at the
location of our observations, and it is not considered further. The distance to \gal\ is somewhat uncertain,
with values reported in the literature ranging from $\sim 4-23~\rm Mpc$. 
We follow \citet{uso03} and \citet{ada11} in this work and assume a distance of $D_{\rm gal} = 10~\rm Mpc$, with  
a corresponding angular scale of $\alpha = 48.5~\rm pc/arcsec$. We note, however, that our results are
based on distance-independent quantities, such as SB and relative separations in the plane of the sky.

\subsection{ESO data product}

For the preparation of the first dataset (hereafter the ESO data cube), we perform sky subtraction on the individual
exposures using the {\tt muse\_scipost} recipe provided within the ESO pipeline. This procedure subtracts a
sky model from the data, correcting for local variation of the line spread function (LSF) in an attempt to minimise
the residuals of bright sky lines. The sky continuum level is computed internally, by selecting a range
of pixels with low illumination to avoid the presence of sources. 

Following sky subtraction, we align all the exposures relative to each other by using continuum-detected sources
as reference. Subsequently, we reconstruct a final data cube using the {\tt muse\_exp\_combine} recipe
that resamples data on a regular cube of $1.25~$\AA\ in the spectral directions, and $0.2~\rm arcsec$
in the two spatial directions. As a last step, we correct the absolute astrometric solution using
the Sloan Digital Sky Survey (SDSS) imaging as reference system \citep{sdss}. We further test the
quality of the flux calibration against SDSS using galaxies in the field
finding good agreement (within $\sim 15\%$).

Inspection of the final data cube reveals the presence of sky residuals with amplitude comparable to the signal we wish to
detect. For this reason, we will only use the ESO product as a reference grid for computing
the astrometric and wavelength solution during the reconstruction of new data cubes that are post-processed with additional
software, as described in the following sections. 

\subsection{\cubex\ data product}\label{sub:cubex}

The second dataset (hereafter the \cubex\ data cube)
is prepared using a combination of procedures distributed as part of the \cubex\ package
(Cantalupo in prep.). At first,  we reconstruct a resampled data cube for each exposure, after it
has been processed for basic calibrations using the ESO pipeline. At this stage, sky subtraction has not
been performed, and we use the ESO data cube as a common reference frame for the final astrometric solution
of individual exposures. All the subsequent post-processing techniques are applied to these reconstructed data cubes,
thus avoiding multiple interpolations of the data. 

\begin{figure}
  \centering
  \includegraphics[scale=0.42]{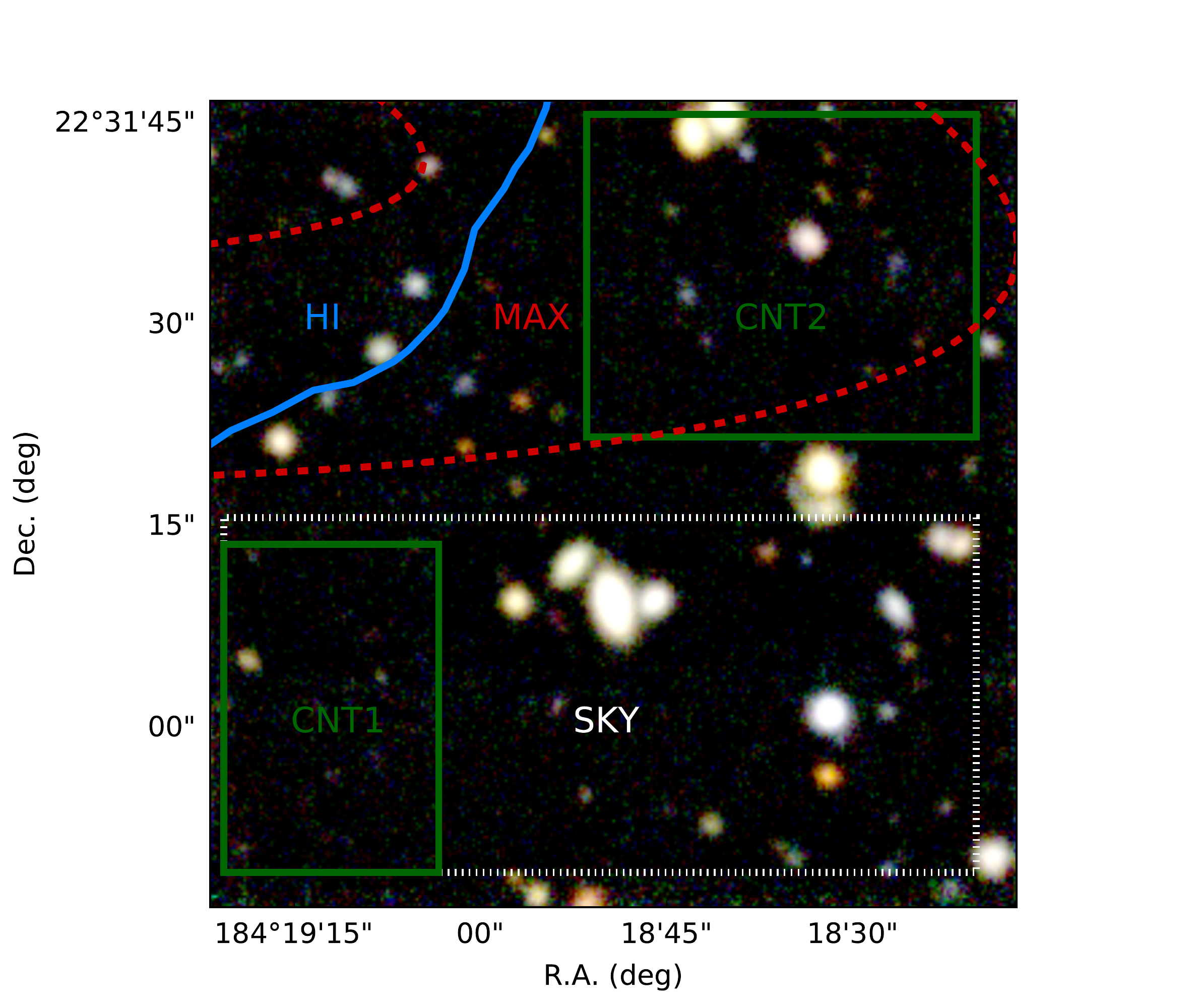}
  \caption{The MUSE field of view shown in a false colour image obtained from three 1000~\AA\ wide images reconstructed 
    from the \cubex\ data cube. The location of five regions that contain pixels used to generate deep stack spectra
    throughout our analysis are also displayed. These regions are defined in Sect. \ref{sub:pixreg}.}\label{fig:musefov}
\end{figure}

The next step uses the {\tt CubeFix} procedure to 
minimise the residual illumination differences that are not fully corrected by flat fields across the 24
integral field units (IFUs) which compose the MUSE instrument. This correction is achieved by using both the sky
lines and the continuum sky emission to rescale slices\footnote{In MUSE, a slice is the basic unit inside an IFU,
  and corresponds to a $0.2\times15~\rm arcsec^2$ segment in the spatial direction.}
relative to each others, also accounting for wavelength-dependent variations. This step ensures that residual
differences in illumination across the field are removed, achieving a uniformity of better than
$\sim 0.1\%$ of the sky level on average \citep{bor16}.
After this correction, the sky is subtracted from the resampled cubes
using the flux-conserving {\tt CubeSharp} procedure, which is
designed to minimise the residuals arising from variations
in the LSF across different IFUs.
The above steps are iterated twice using the products of the previous iteration to identify and
mask astronomical sources within the cube. To minimise the risk of altering the astrophysical signal during sky subtraction,
when computing the normalisation of the sky flux as a function of wavelength with {\tt CubeSharp}, 
we further mask the top half of the MUSE FOV, where the H$\alpha$ signal is expected to lie (see Figure \ref{fig:musefov}).
In the end, a final data cube is reconstructed by averaging individual exposures applying
a 3$\sigma$-clipping algorithm.

\subsection{\zapline\ data product}\label{sub:zap}

The preparation of the third dataset (hereafter the \zapline\ data cube) follows a procedure similar to the one adopted
for the \cubex\ data cube, but using Python code we developed to perform the illumination correction and
the \zapline\ package \citep{sot16} to perform sky subtraction. 

As done previously, we resample each exposure after basic reduction with the ESO pipeline
onto a regular grid, using the ESO data cube as reference for the astrometry and wavelength grid.
At this stage, we also produce masks that trace each
voxel in the reconstructed cube back to the original MUSE IFU, retaining also information on the pixel location within
stacks\footnote{Within MUSE, a stack is a  group of 12 slices within an IFU. A MUSE IFU contains 4 stacks of 12 slices each.
A voxel is defined as a datapoint inside a cube, while a spaxel is a pixel in the spatial direction.}.  
After masking continuum-detected sources, we use sky regions to map and correct the
residual illumination differences first across IFUs as a function of wavelength using coarse spectral bins of 100~\AA,
and then across stacks collapsing the entire cube into an image.
These corrections are of the order of $\lesssim 1\%$ and, by construction, they preserve the mean flux across
the cube as a function of wavelength.  We have verified that the photometric properties of sources
detected across the field are preserved when compared to the ESO data cube. 
A major difference with {\tt CubeFix} is that we do not
correct slices individually and we do not separate the contribution of sky lines and sky continuum when computing
the scaling factors.

After this step, we use the \zapline\ code to perform sky subtraction.
As described in \citet{sot16}, \zapline\ employs principal component analysis (PCA) to describe and subtract the sky emission
within each MUSE voxel. As for the \cubex\ product, we reduce the risk of subtracting astronomical signal
by applying a mask in addition to the internal \zapline\ algorithms that minimise the inclusion of pixels with sources
in the computation of the PCA components. To this end, 
we compute the sky eigenspectra using only pixels in the bottom half of the MUSE FOV, in a ``SKY region'' (see Sect. \ref{sub:pixreg} and Figure \ref{fig:musefov}) that does not overlap with the region where
\ha\ is  maximal in the model by \citet{ada11}.
Finally, we combine all the exposures into a mean data cube.

\subsection{Definition of pixel regions}\label{sub:pixreg}

Throughout our analysis, we make extensive use of regions in the image plane
to generate deep composite spectra. These regions are
shown in Figure \ref{fig:musefov}, superimposed on a false colour image of the
MUSE FOV that we obtain from three 1000~\AA\ wide images extracted from the \cubex\ data cube.

These regions are defined as it follows. The first region, labelled ``\HI\'', contains all the
pixels within the \HI\ column-density contour $N_{\rm HI} = 10^{19}~\rm cm^{-2}$ 
that is marked by the blue solid line. The second region, labelled ``MAX'', is enclosed by the
two red dashed contours where the \citet{ada11} model forecasts
maximal SB from gas photoionized by the UVB. The third ``SKY'' region, enclosed by white dotted lines,
encompasses pixels far from the region where the SB is expected to be maximal. Finally, we define two
control regions (``CNT1'' and ``CNT2'') that will be used  to test the
quality of the sky subtraction and for the preparation of mock data cubes as described in the following section.

Throughout our analysis, we exclude pixels at the position of sources detected
via continuum emission, a task that is easily achieved thanks to the excellent image quality of MUSE.
To this end, we run {\sc SExtractor} \citep{ber96} on a deep white image that we obtain by collapsing the
data cube along the wavelength axis. For this, we choose the \cubex\ data product as it
provides the best image quality given that the illumination correction is performed
at the slice level. To ensure that the full extent of the sources are masked down to faint SB levels, we
produce a segmentation map using a low detection threshold, equal to the sky root-mean-square (RMS).
To avoid the inclusion of spurious sources, the minimum area for source detection is set to 15 spaxels, corresponding to
objects of $\gtrsim 0.9~\rm arcsec$ in diameter. Visual inspection confirms that the segmentation map is
successful in masking all the sources where continuum emission is seen in the deep white image.

\subsection{Preparation of mock data cubes}\label{sub:mock}

To better understand the performance of the adopted reduction techniques, and to assess the presence of
systematic errors throughout our analysis, we make use of mock data cubes that contain
emission lines injected at wavelengths and positions chosen as described below.   

All mock emission lines have Gaussian profiles with a full-width at half-maximum of
2.6~\AA\ that matches the resolving power of MUSE at the wavelength
of interest, $R \sim 2550$ at $\lambda \sim 6574~$\AA.
As discussed below, this is the wavelength at which \ha\ recombination is expected given the radial
velocity of \gal.
Mock lines are generated at three different wavelengths  (see Figure \ref{fig:1dspec} for examples
of observed spectra) chosen in the following way.
First, we create a line with SB\footnote{Throughout this work, we will make use of the symbol $\mu$
  to identify the line SB, and $\mu_{20}$ to identify the line SB in units of $10^{-20}~$\sbline.
  Similarly, we will indicate the continuum SB with the symbol $\mu_{\rm c}$ and use  $\mu_{\rm 20,c}$
  for values in units of $10^{-20}~$\sbcgs.}
$\mu = 2\times 10^{-19}~$\sbline\ at $\lambda = 6574~$\AA. This choice allows us to test whether emission at this wavelength can be recovered correctly in our analysis.
This signal is injected in pixels within the CNT1 region, at a location where no signal is expected.
Next, we create a line with $\mu = 2\times 10^{-19}~$\sbline\ at $\lambda = 6550~$\AA, which is adjacent to a
bright sky line at $\lambda = 6553~$\AA. This mock line is injected  in both the CNT1 and CNT2 region, and 
it enables tests for the presence of any bias when measuring signal in the wings of bright sky lines.
Finally, we create a line with $\mu = 3\times 10^{-19}~$\sbline\ at $\lambda = 6590~$\AA,
a wavelength free from bright sky lines. This line is injected both in the CNT1 and CNT2 regions and
it is used as a baseline calibration to test whether our procedures are flux conserving.

These mock lines are injected within individual exposures, by adding flux to the data cubes that have
been resampled on the final ESO data cube after performing basic calibrations only. These individual exposures are
then processed as described in Sect.~ \ref{sub:cubex} and Sect.~\ref{sub:zap}using both the \cubex\ and \zapline\ pipelines.

\begin{figure*}
  \begin{tabular}{c}
    \includegraphics[scale=0.45]{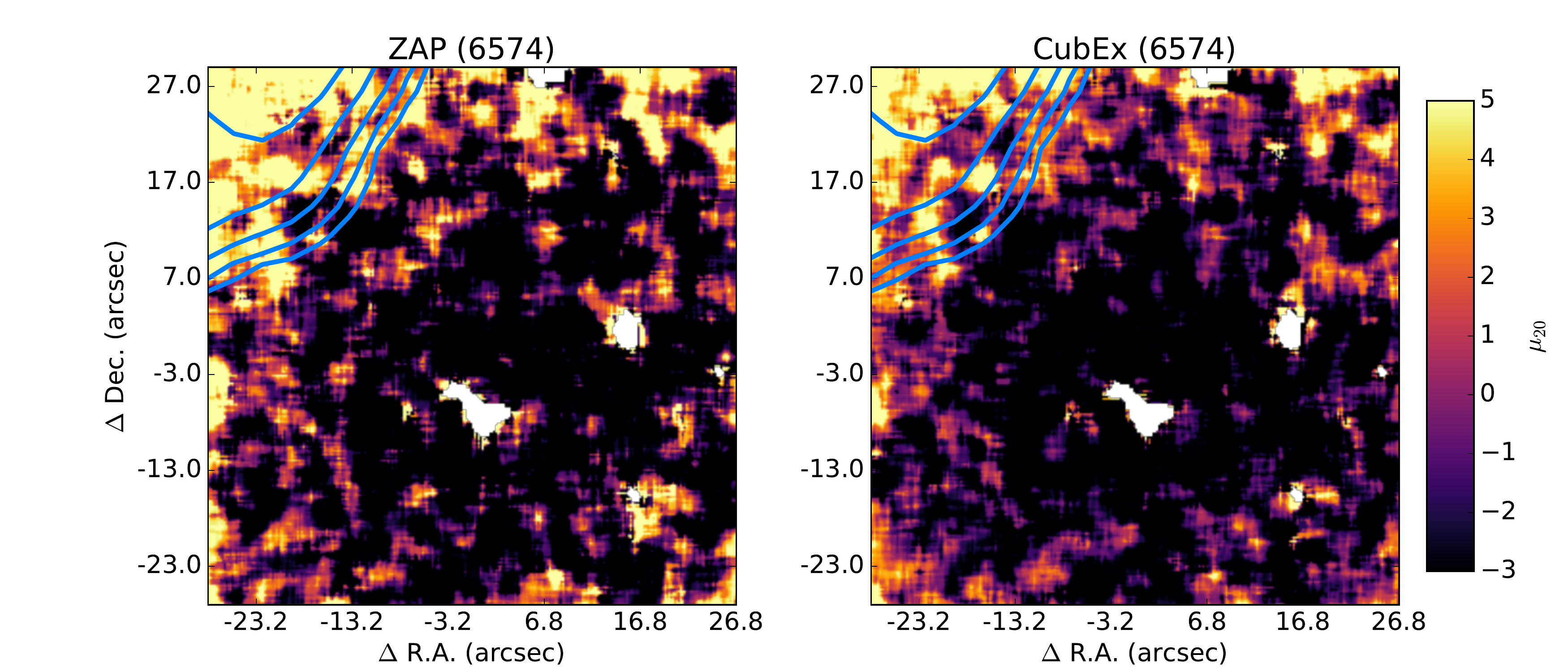}\\
    \includegraphics[scale=0.45]{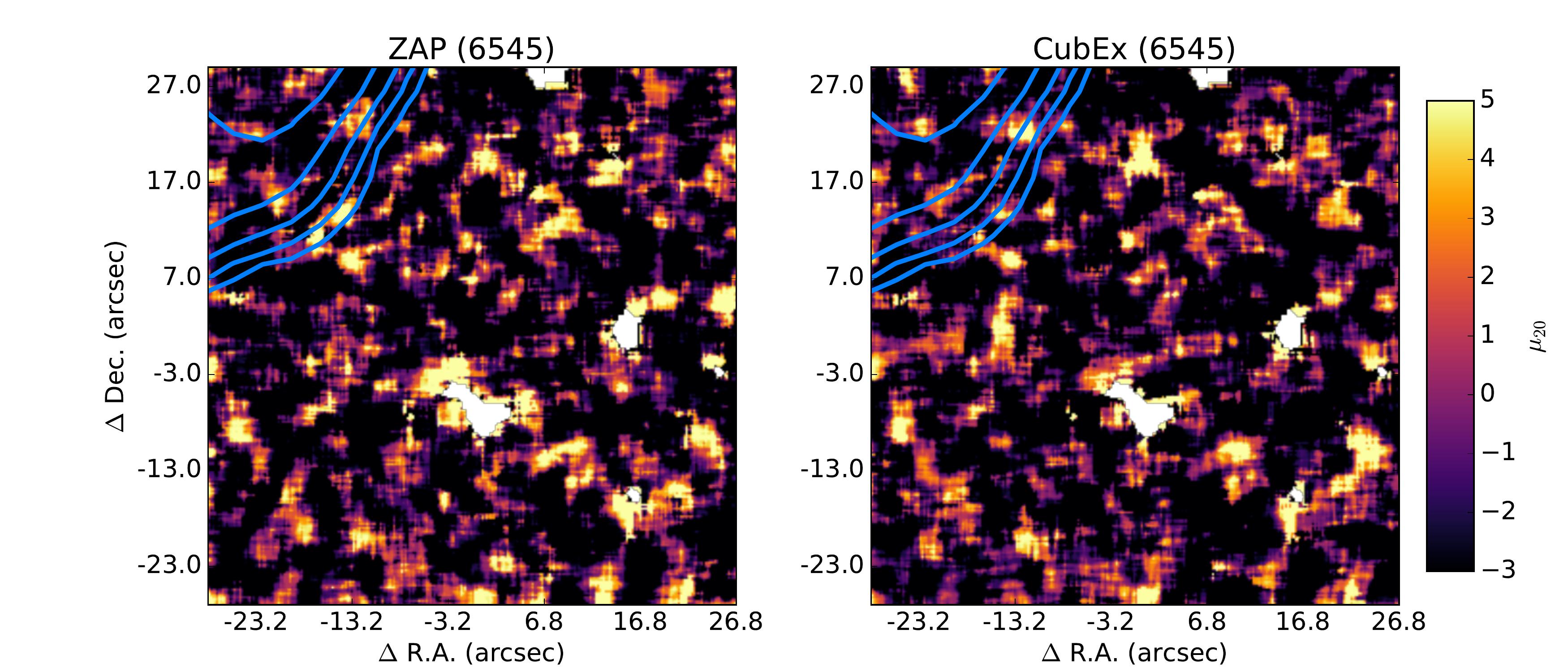}
  \end{tabular}
  \caption{Optimally-weighted SB maps extracted at the wavelength of the expected signal ($\lambda = 6574~$\AA, top),
    and in a control region (bottom) centred at $\lambda = 6545~$\AA, which is next to a bright sky line
    at $\lambda \sim 6553~$\AA. North is up and East is to the left.
    The left- and right-hand panels show maps from the \zapline\ and \cubex\ data cube, respectively.
    Data have been smoothed by a median filter with size of 1.2 arcsec.
    \HI\ contours are overlaid in blue, and white patches are sources that have been masked.
    Extended \ha\ emission overlapping with the \HI\ column density $N_{\rm HI} \ge 10^{19}~\rm cm^{-2}$ is
    visible at $\lambda \sim 6574~$\AA\ in both the \cubex\ and \zapline\ cubes. The lack of emission in the control regions
    rules out a spurious origin for this signal due to artefacts such as scattered light.}\label{fig:sbmaps}
\end{figure*}

\section{Analysis of MUSE observations}\label{sec:data}

\subsection{Theoretical expectations for \ha\ emission near \gal}\label{sec:theo}

The wavelength at which \ha\ recombination due to the ionising UVB is expected
can be computed given the heliocentric radial velocity of \gal,
$v_{\rm HI} = 406.8 \pm 0.1 ~\rm km~s^{-1}$ \citep{uso03}, and the galaxy rotation curve
known from 21cm observations.  At the position of our observations, $v_{\rm H\alpha} \sim 510~\rm km~s^{-1}$
(or $\lambda_{\rm H\alpha} = 6574~$\AA) with an uncertainty of 0.5~\AA\ \citep{uso03,ada11}.
The emission line is expected to be unresolved at the moderate resolution of MUSE
($R\sim 2550$ at these wavelengths), but as in \citet{ada11}, we assume a conservative
window of $\pm 100~\rm km~s^{-1}$ ($\pm 2.2~$\AA) over which the line can be detected due to
variations in the gas velocity field across the MUSE FOV. Under the general assumption that the gas
is in photoionization equilibrium, the emission line is further expected to have an order of magnitude SB of
$\mu_{20} \sim 10$, with the exact normalisation depending on the UVB photoionization rate and spatial
location, as described in Sect. \ref{sec:uvb}. According to the model by \citet{ada11},
the emission is further expected to be maximal within the region labelled \HI\ in Figure \ref{fig:musefov}.
However, we will present new models that supersede this prediction in Sect. \ref{sec:uvb}.

\subsection{Searching for \ha\ recombination in MUSE data}

\subsubsection{Analysis of two-dimensional maps}

To visually evaluate if any signal is detected in MUSE data at the expected position,
we extract SB maps from the \zapline\ and \cubex\ data products
by slicing the cubes in a window centred at $\lambda_{\rm H\alpha} = 6574~$\AA\ (Figure \ref{fig:sbmaps}; top panels).
To maximise the signal-to-noise ($\rm S/N$) of these maps, we compute the mean SB by summing flux along the wavelength
direction while weighting according to a normalised Gaussian of width $\sigma = 1.095$\AA, which is
matched to the MUSE resolution at these wavelengths.
As discussed below, this procedure does not provide the best estimate for the total line flux, but it is suitable for
a visual exploration of the presence of signal within maps of maximal $\rm S/N$.

After applying a two-dimensional (2D) median filter of width 1.2 arcsec, visual
inspection of these maps reveals the presence of extended emission
in the North-East (top left) corner of the FOV in both data cubes, inside the region overlapping with the 21cm detection. 
Conversely, no extended emission is found in the top right part of the FOV, within the region where the
SB is expected to be maximal in the model by \citet{ada11}, perhaps with the exception of the edge of the map. Furthermore,
the presence of positive fluctuation at the outskirts of the map can be noted in the South-East direction, indicating
that artefacts may be present at the edge of the FOV.

Relying again on visual inspection, we test whether the signal visible in these maps
can be attributed to scattered light in one of the MUSE corners. Having produced the final data cubes by averaging exposures at
different position angles, there should be no preferential direction in the final combined cubes and, in principle,
any residual illumination pattern should not appear in a single corner. Nevertheless, we explicitly check for the presence
of spurious scattered light as well as astrophysical signal with a broadband spectrum by extracting SB maps centred at
$\lambda = 6545$~\AA, that is only $\sim 30$~\AA\ from the region where we expect the \ha\ emission line. Inspection of these
maps (Figure \ref{fig:sbmaps}; bottom panels) does not reveal a prominent positive flux,
thus excluding ``white'' light as the origin of the positive signal at $\lambda = 6574~$\AA.
Similarly, no flux excess is visible in maps extracted in a window centred at $\lambda = 6590$~\AA\ (not shown),
ruling out spurious signals with a broad spectrum.

\begin{figure*}
  \centering
  \includegraphics[scale=0.4]{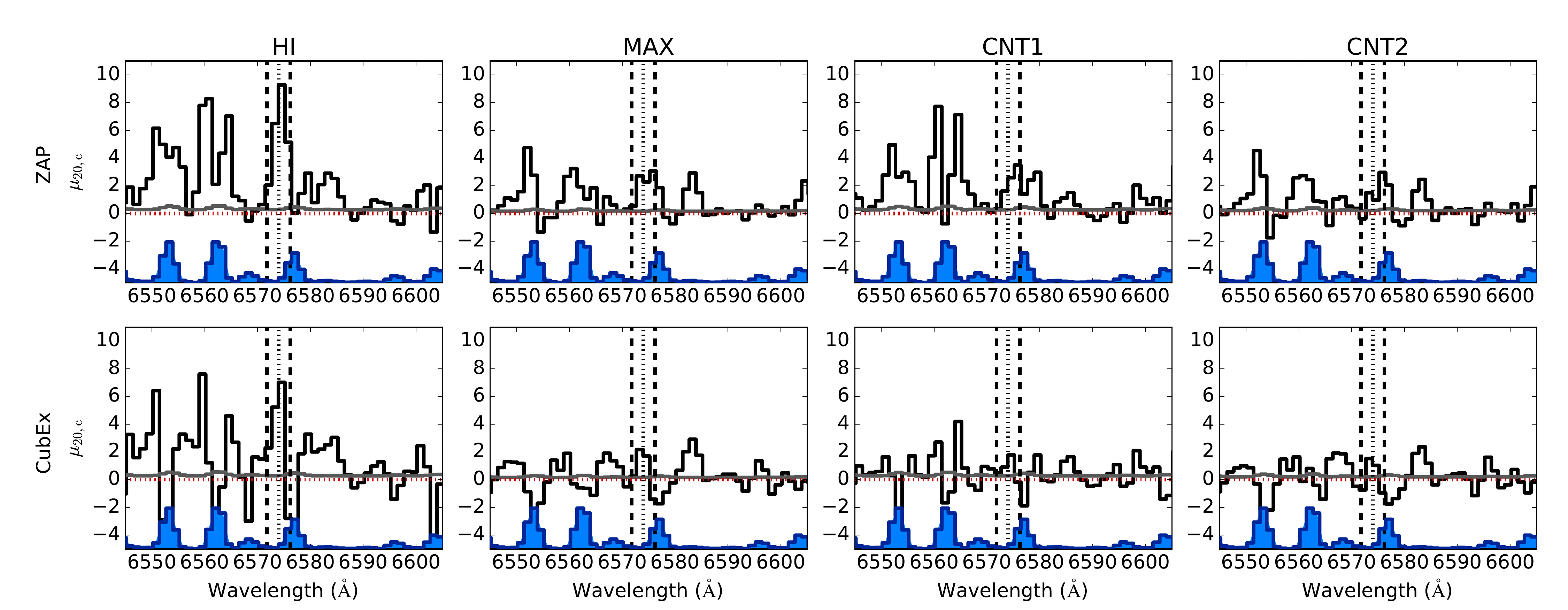}\\
  \caption{Gallery of mean spectra (black lines) obtained combining pixels inside the regions 
    defined in Figure \ref{fig:musefov}, as labelled at the top of each panel. The associated statistical errors are shown in grey,
    and the red dotted lines mark the zero flux level. The blue filled spectra at the bottom of each panel
    show the mean sky SB computed from the same data. For visualisation purposes, the sky flux has been
    scaled by a factor of $1/600$ and offset to $\mu_{20} = -5$.
    The top row shows results from the \zapline\ reduction, while the bottom row shows spectra from
    the \cubex\ reduction. The wavelength window within which the \ha\ emission line is expected
    is marked by vertical dashed lines, while the vertical dotted lines mark the expected position for the
    centroid of the emission line. A signal consistent with the expected \ha\ recombination line from \gal\
    is detected inside the \HI\ region, both in the \zapline\ and \cubex\ reduction.}\label{fig:1dspec}
\end{figure*}

\subsubsection{Analysis of the mean spectra}

Having established that data reveal a positive flux  that is not associated with
signal (astrophysical or instrumental) across a broad wavelength interval (with $\Delta \lambda > 40~$\AA),
we next characterise the spectral properties of the MUSE cubes by constructing mean 
spectra by averaging flux from all of the pixels inside the regions defined in Sect. \ref{sub:pixreg}.
During this step, we exclude pixels that overlap with the position of continuum-detected sources. We also
characterise the error on the mean by propagating the variance computed during data reduction, and also by means
of empirical measurements of the RMS in each wavelength layer. The two methods are found to yield comparable
error estimates. Finally, we ensure that flux in the regions free from sky lines
in the wavelength interval $\lambda = 6483-6493~$\AA\ and $\lambda = 6609-6623~$\AA\ averages to zero,
by subtracting a constant of $\lesssim 2\times 10^{-20}~$\sbcgs\ for the \zapline\
reduction and of $\lesssim 0.5\times 10^{-20}~$\sbcgs\ for the \cubex\ reduction.
A gallery of the mean spectra constructed for the \zapline\ and \cubex\ cubes inside different apertures
is in Figure \ref{fig:1dspec}.

Additional properties of the signal seen in Figure  \ref{fig:sbmaps} can be inferred by inspecting the mean spectra
from different apertures. First of all, in agreement with the results derived from the optimally-extracted 2D maps,
an emission line is detected at $\lambda \sim 6574~$\AA\ within the \HI\ region (left-hand panels).
The significance of this detection exceeds $\sim 10\sigma$ based on
statistical errors, here defined as the photon and detector noise that are estimated by propagating the variances
of these contributions through the calibration and reduction procedures.
However, the systematic errors are the dominant source of uncertainty, which
arises from imperfect calibrations and sky residuals that are not
fully corrected by the above procedures.
As described below, we characterise these additional errors by means of 
mock data and by comparing measurements using different reduction techniques and different subsets of the data.

Figure \ref{fig:1dspec}
reveals that the emission line is detected at the wavelength
where \ha\ recombination from \gal\ is expected, it is visible both in the \zapline\ and \cubex\ cubes, and 
it appears in the region enclosed by the \HI\ contour with $N_{\rm HI} = 10^{19}~\rm cm^{-2}$.
Combined, these three pieces of evidence suggest that the
signal seen in Figure \ref{fig:sbmaps} and Figure \ref{fig:1dspec} is real and that it is consistent with
\ha\ emission from the outskirts of \gal, as expected for gas  photoionized by the UVB.
Finally, no prominent emission is visible inside the CNT1 and the CNT2 regions
(third and fourth panels in Figure \ref{fig:1dspec}), indicating that the positive flux fluctuations that are
visible at the West and North edge of the FOV in Figure \ref{fig:sbmaps} are not related to
artefacts that can mimic an emission line at the wavelength expected for \ha\ recombination in \gal.

Moreover, no prominent signal is detected when inspecting the mean spectra in the MAX region (second panels from the left),
in agreement with what was found in the SB maps.
The lack of strong signal inside the region that was predicted to contain the strongest emission
in the model by \citet{ada11} may appear puzzling at first. However, as we discuss in detail in Sect. \ref{sec:uvb},
this is fully consistent with our revised photoionization model for \gal.
This discrepancy can be explained by the fact that the model
by \citet{ada11} overestimated the extent of the \HI\ profile, predicting a more extended SB profile than is
warranted by current data. This unfortunate mismatch between the model by \citet{ada11}
and the data has led us to focus our MUSE observations
within a region of lower SB, with the brighter emission being confined in the corner of the FOV, in a region
that overlaps with the \HI\ emission detected at 21 cm. 

Focusing again on the feature at $\lambda \sim 6574~$\AA, 
it is evident from the spectra shown in Figure \ref{fig:1dspec}
that random errors are negligible compared to the systematic uncertainty arising from residuals of bright sky lines.
Indeed, the wavelength of the expected signal at  $\lambda_{\rm H\alpha} = 6574~$\AA\ falls at just $\sim 3~$\AA\ blueward
of the $\lambda \sim 6577~$\AA\ sky line, causing a partial blend at the resolution of MUSE.
Furthermore, comparisons of the mean spectra in different panels reveal that the quality of the
sky subtraction is lower in the North-East corner of the FOV (left-hand panels)
compared to what can be achieved in the central parts of the detector (MAX and control regions, right-hand panels).
We speculate that this effect is due to small errors in the geometric distortion correction and wavelength
calibrations at the very edge of the FOV.

\begin{figure}
  \centering
  \includegraphics[scale=0.35]{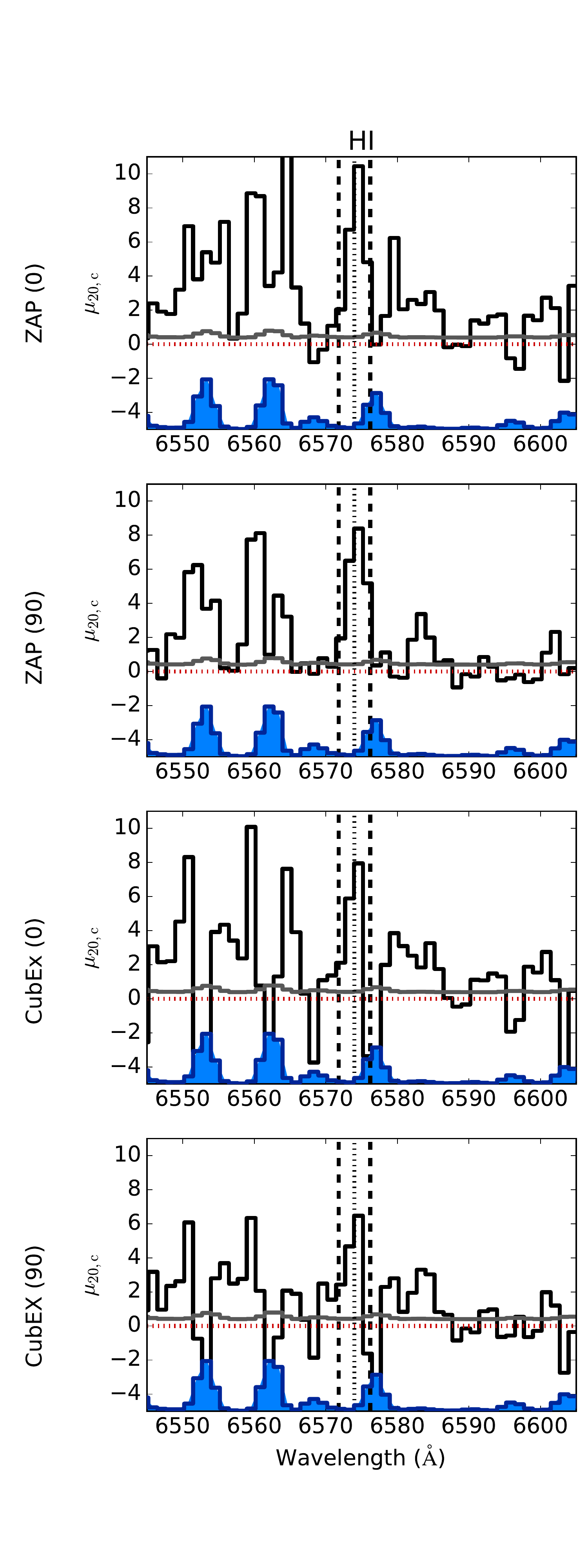}\\
  \caption{Same as Figure \ref{fig:1dspec}, but for the mean spectra extracted in the \HI\ region using
    independent subsets of exposures obtained at position angles of $0/180$ degrees (labelled 0) and
    $90/270$ degrees (labelled 90). Spectra from \zapline\ cubes are shown in the top two panels, while spectra from
    \cubex\ cubes are in the bottom two panels. A line at $\lambda \sim 6574~\rm \AA$ is consistently
    detected in independent sets of exposures.}\label{fig:1dspec_rot}
\end{figure}

\begin{figure}
  \centering
  \includegraphics[scale=0.32]{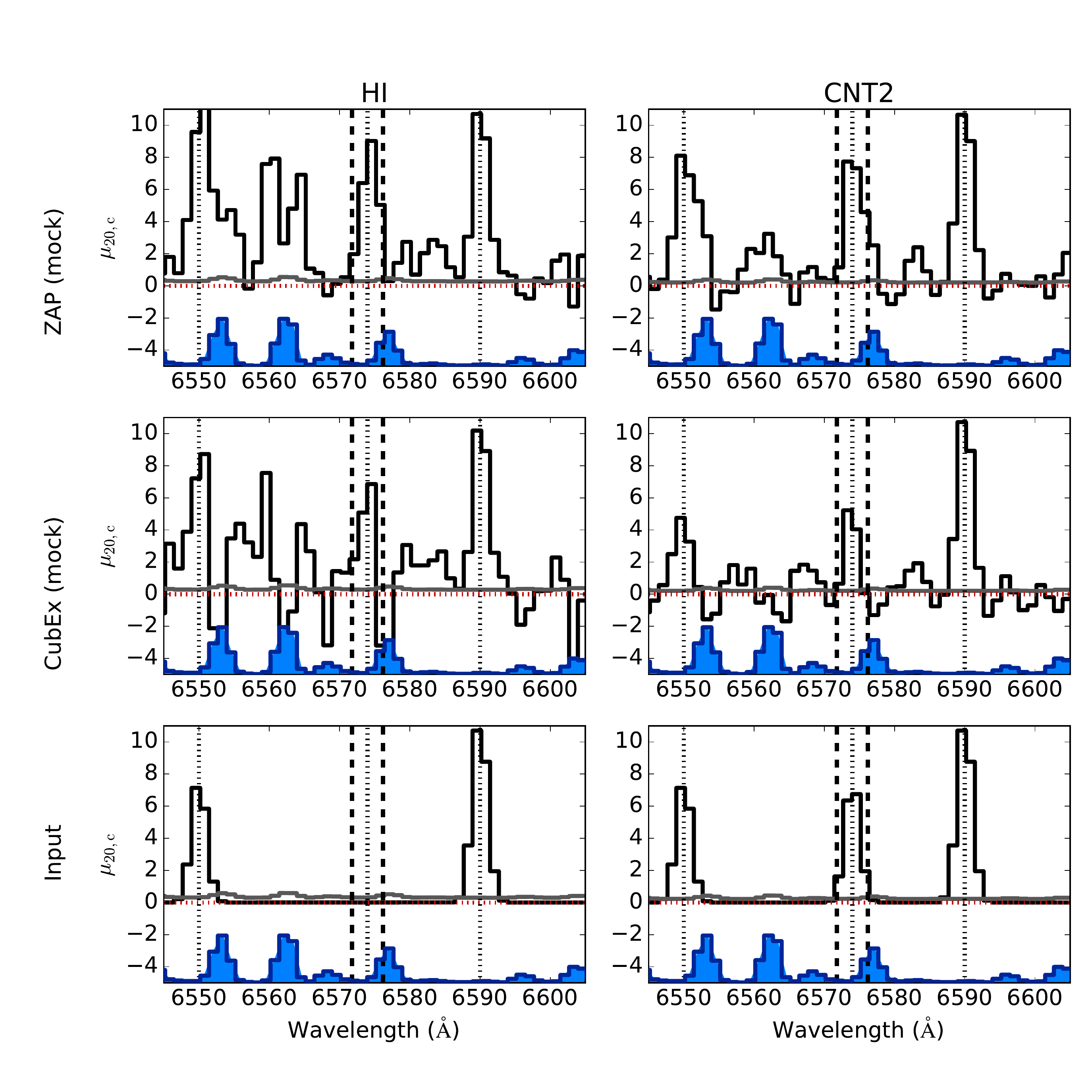}\\
  \caption{Same as Figure \ref{fig:1dspec}, but for mean spectra obtained after injecting mock
    lines as described in Sect. \ref{sub:mock}, the position of which is marked by vertical dotted lines.
    The input mock spectra are shown in the bottom
    panels, while the top and middle panels show the recovered SB for the \zapline\ and \cubex\ reductions,
    respectively. Despite noticeable contamination arising from the wings of sky lines, the mock emission lines are
    recovered at the expected locations.}\label{fig:1dspec_mock}
\end{figure}

\subsubsection{Additional tests on the origin of the detected signal}

Given the presence of prominent residuals next to sky lines, we should take particular care when interpreting the origin of this signal and, most importantly, when quoting the significance of our measurement. To corroborate our earlier conclusion
that the line detected is H$\alpha$ from \gal, we  perform three additional tests.

Firstly, we  compare the shape of the $\lambda \sim 6574~$\AA\ line with the shape of the residuals associated
with the subtraction of the bright sky lines at $\lambda \sim 6553~$\AA\ and $\lambda \sim 6562~$\AA\ (Figure \ref{fig:1dspec}).
Residuals in the \zapline\ reduction, albeit strong,
appear to span the entire width of these sky lines. Conversely, the line at $\lambda \sim 6574~$\AA\
is offset from, and does not overlap with, the wavelength position of the sky residual associated with the
$\lambda \sim 6577~$\AA\ sky line. A similar behaviour is visible in the \cubex\ reduction, with positive and negative
residuals aligned with the sky lines at $\lambda \sim 6553~$\AA\ and $\lambda \sim 6562~$\AA,
and a clearly asymmetric profile next to the $\lambda \sim 6577~$\AA\ sky line. Furthermore, it should be noted that
the emission line at $\lambda \sim 6574~$\AA\ is the strongest feature in these spectra,
despite the sky line at $\lambda \sim 6577~$\AA\ being the faintest in this wavelength interval.

As a second test to corroborate the detection of a line at $\lambda \sim 6574~$\AA,
we analyse independent subsets of exposures as shown in  Figure \ref{fig:1dspec_rot}.
To this end, we generate mean spectra in the \HI\ region after reconstructing two independent
data cubes, using half of the total number of exposures. To simultaneously test for
subtle systematic errors associated with the instrument rotation, we group exposures according to the instrument
position angle. In Figure \ref{fig:1dspec_rot}, we group the 90 and 270 degree rotation in what
we label the ``90 set'', and the 0 and 180 degree rotation in what we label the ``0 set''.
The resulting spectra reveal that, while the shape of the residuals associated with the $\lambda \sim 6553~$\AA\
and $\lambda \sim 6562~$\AA\ sky lines change with the instrument position angle in both the \cubex\ and \zapline\ reductions,
the line at $\lambda \sim 6574~$\AA\ is consistently recovered with a similar shape, as expected for an
astrophysical signal.

As a third and final test, we make use of the mock cubes described in Sect. \ref{sub:mock}.
Using mock data, we check explicitly whether the shape of the sky residuals discussed above is a trustworthy indicator
of the presence of a line at $\lambda \sim 6574~$\AA.
For this, we inject mock lines with the spectral properties shown in the bottom panels of Figure \ref{fig:1dspec_mock}
and we perform illumination corrections and sky subtraction using
the same pipelines used for the data.
Particularly relevant for our test is the fact that the mock line at $\lambda \sim 6550~$\AA\ is
offset by $\sim 3~$\AA\ from a sky line, as is the case for the line at $\lambda \sim 6574~$\AA\ compared to the
$\lambda \sim 6577~$\AA\ sky line.  Focusing on the \HI\ region first (left-hand panels of Figure \ref{fig:1dspec_mock}),
it is evident that the mock line at $\lambda \sim 6550~$\AA\ is recovered by our analysis. The spectra also exhibit
a residual with a prominent excess at bluer wavelengths,
similar to the feature at $\lambda \sim 6574~$\AA\ where the astrophysical
signal is expected and no mock line is injected.

The right-hand panels of Figure \ref{fig:1dspec_mock} show  how mock emission lines, including the mock signal at
$\lambda \sim 6574~$\AA, are recovered inside the CNT2 region in both the
\zapline\ and \cubex\ reductions. This implies that our reduction procedures are flux conserving,
and that the lack of appreciable emission in the MAX region (Figure \ref{fig:1dspec})
is genuine and not attributable to improper sky subtraction.
Most notably, the mock line at $\lambda \sim 6590~$\AA\ that is
far from sky lines is recovered with very high precision and accuracy,
implying that MUSE is potentially well suited for measurements with $\lesssim 10\%$ error. We will return
to this point in Sect. \ref{sec:end}.

In summary,  we have shown that: i) a feature is detected at $\lambda \sim 6574~$\AA\ and it
is consistently present in two independent data reductions and in two independent sets of exposures with different
instrument rotations; ii) the recovered line has a profile consistent with real signal close to a sky line; iii) the emission
overlaps with the location where \HI\ is detected with $N_{\rm HI} \gtrsim 10^{19}~\rm cm^{-2}$.
Altogether, these pieces of evidence corroborate the detection of
an extended low SB signal that is consistent with our expectation of \ha\ emission from
gas that recombines following photoionization from the UVB at the edge of \gal. 

\subsection{Measurement of the detected emission line}

In the previous section, we have shown how data support the detection of \ha\ emission in the outskirts of \gal.
However, our analysis has also demonstrated that strong residuals associated with sky lines
are present, and that they dominate the error budget in our measurement. In this section, we attempt to quantify the
amplitude of this systematic uncertainty.

Starting with the analysis of the 1D spectra shown in Figure \ref{fig:1dspec}, we integrate the SB within
a $\pm 2.2~$\AA\ window (as justified in Sect. \ref{sec:theo}) around the
wavelength $\lambda_{\rm H\alpha} = 6574~$\AA, finding $(1.4\pm 0.1)\times 10^{-19}~$\sbline\ for the \cubex\
reduction and $(2.6\pm 0.1)\times 10^{-19}~$\sbline\ for the \zapline\ reduction.
Here, the uncertainty quoted for individual measurements reflects only the statistical error. 
The reason for the different SB values is attributable to the fact that the \cubex\ reduction
appears to better suppress the sky line residuals when compared with the \zapline\ reduction.
This effect can be quantified using mock data. Indeed, for an input mock line of $2\times 10^{-19}~$\sbline\ at
$\lambda = 6550~$\AA, we recover an integrated signal
of $(2.2 \pm 0.1) \times 10^{-19}~$\sbline\ from the \cubex\ reduction.
Conversely, the \zapline\ reduction yields an integrated SB of $(3.5 \pm 0.1) \times 10^{-19}$\sbline,
revealing that positive residuals of the order of $\sim 1 \times 10^{-19}~$\sbline\ are present next to sky lines.
Far from the wings of the sky lines (e.g. at $\lambda \sim 6590~$\AA), both reduction techniques are
able to recover the input line SB to within the associated statistical errors of $\sim 5\%$.
Based on this analysis, in the following we assume that the \cubex\ reduction yields a more accurate value for the line SB.

To estimate the amplitude of the systematic uncertainty, we proceed as follows.
First, we measure the SB values for the two independent cubes which we obtain by combining independent
sets of data, finding consistent values of $(1.5 \pm 0.1) \times 10^{-19}~$\sbline\ and
$(1.4 \pm 0.1) \times 10^{-19}~$\sbline\ for the 0 and 90 sets, respectively.
This finding rules out the presence of systematic differences associated with the instrument rotation.
However, by integrating the line SB within the control regions defined above both at $\lambda = 6550~$\AA\ and
$\lambda = 6574~$\AA, we find fluctuations which are up to one order of magnitude higher than the quoted variance
based on statistical uncertainties.
By comparing  measurements in these control apertures we find a dispersion of $\sim 5\times 10^{-20}~$\sbline,
which we consider a more realistic estimate of the uncertainty of our measurement. 

Finally, we perform two additional tests. First, we perform the wavelength integral on the \cubex\ reduction
by first collapsing the cube along the wavelength direction and then adding the SB in pixels within the
regions defined in Figure \ref{fig:musefov}. In this case, we find a line SB of
$(1.1 \pm 0.1) \times 10^{-19}~$\sbline, again with a scatter of $\sim 3-4\times 10^{-20}~$\sbline\
within the control apertures. Also in this case, the analysis of the two independent rotations yields consistent results.  
Next, we perform the integration by considering a larger wavelength window of $\lambda = 6569-6581~$\AA,
chosen to encompass the sky line at $\lambda \sim 6577~$\AA.
This choice is dictated by the fact that, by construction, the \cubex\ reduction is flux conserving across
wavelength windows that are larger than the widths of the sky lines.
In agreement with the previous measurements, we find a value of $(1.1 \pm 0.1) \times 10^{-19}~$\sbline.

In summary, by analysing both the 1D spectra and the 2D line maps at the position expected for \ha\ recombination due
to the ionisation from the UVB in our updated models for \gal\ (see Sect. \ref{sec:uvb}), we find consistent indications
of the presence of a line with SB $(1.2 \pm 0.1 \pm 0.5) \times 10^{-19}~$\sbline. Here, the first error indicates the
statistical uncertainty and the second error characterises the presence of an additional systematic uncertainty in proximity
to the bright sky line at $\lambda \sim 6577~$\AA\ and at the edges of the FOV.
This value is fully consistent with the detection reported by Uson et al.\footnote{See \url{http://iactalks.iac.es/talks/view/393}} of $(0.96 \pm 0.14) \times 10^{-19}~$\sbline.

\section{Constraints on the UVB intensity}\label{sec:uvb}

After presenting an overview of our new radiative transfer calculations in Sect. \ref{sec:uvbcode},
in Sect. \ref{sec:uvblimit} we describe the procedure adopted to constrain the \HI\ photoionization
rate ($\Gamma_{\rm HI}$) starting from the observed \ha\ SB.

\subsection{Photoionization modelling of \gal}\label{sec:uvbcode}

To predict the \ha\ SB as a function of the UVB intensity,
we construct a photoionization model for \gal, improving upon the analytic calculations
presented in \citet{ada11}.

\subsubsection{Description of the photoionization code}

We model the hydrogen density of \gal\ as an exponential disk
\begin{equation}
n_{\rm H}(R,\bar z)=n_{\rm H,0}\exp{(-R/h_{\rm R})}\exp{(-|\bar z|/h_{\rm z})},
\end{equation}
where $n_{\rm H}(R,\bar z)$ is the total hydrogen number density in cylindrical coordinates ($R,\bar z$), $n_{\rm H,0}$
defines the central density, while $h_{\rm R}$ and $h_{\rm z}$ are, respectively, the radial scale-length and
vertical scale-height of the disk. For a given external and isotropic UVB, we solve for the vertical ionisation and
temperature of the disk at a fixed radial distance $R$ assuming a two-sided plane parallel geometry.
In this way, we are effectively reducing the
full three-dimensional radiative transfer problem to a series of 1D calculations.
The full structure of the galaxy in terms of temperature and ionisation fraction is thus reconstructed
combining results of calculations with plane-parallel geometries at different $R$. Such approximation
is expected to give results that are accurate to within 20-30\% when
compared with a full three-dimensional calculation \citep[see, e.g.,][]{dov94}.

Details of the adopted radiative transfer scheme are described in \citet{haa12}. Briefly, 
the ionisation and thermal vertical structure is solved iteratively for an input
power-law spectrum with spectral slope $1.8$. Ionisation equilibrium is achieved by balancing
radiative recombinations with photoionization, including the formation and propagation of recombination radiation from
\ion{H}{II}, \ion{He}{II} and \ion{He}{III}. For the thermal structure, photo-heating is balanced by free-free, collisional
ionisation and excitation, and recombinations from \ion{H}{II}, \ion{He}{II}, and \ion{He}{III}.
In our calculation, we assume a number density ratio $\rm He/H = 1/12$.
A current limitation of the model is that we do not include metal lines nor dust.

Once the ionisation and thermal state of the gas are known, we compute the \ha\ emissivity as 
\begin{equation}
\epsilon_{\rm{H}\alpha}(R,\bar z)=h\nu_{\rm{H}\alpha} \alpha^{\rm eff}_{\rm{H}\alpha}(T)\, n_{\rm p}(R,\bar z)n_{\rm e}(R,\bar z)\:,
\end{equation}
where $n_{\rm p}$ and $n_{\rm e}$ are the proton and electron number densities,
and $\alpha^{\rm eff}_{\rm{H}\alpha}$ is the effective case A recombination rate taken from \citet{peq91}.
Finally, we derive the \HI\ column density and the \ha\ SB maps from an
integration along the line-of-sight of the neutral hydrogen number density and \ha\ emissivity.
Specifically, for a given viewing angle $i$, we compute the projected maps of 
$N_{\rm HI}(b_1,b_2)$ and $\mu(b_1,b_2)$, where $b_1$ and $b_2$ describe a new coordinate system along the
semi-major and semi-minor axis of the projected ellipse. The relations connecting the cylindrical coordinate
system $(R,\bar z)$ to the projected position $(b_1,b_2)$ can be easily obtained from the following coordinate
transformations \citep[e.g.][]{ada11}:

\begin{equation}
|\bar z|=|\rho \cos{i} + b_2\sin{i}|
\end{equation}
and
\begin{equation}
R=\sqrt{(\rho\sin{i}-b_2\cos{i})^2+b_1^2}.
\end{equation}
Here $\rho$, which ranges from $-\infty$ to $+\infty$, is the distance from the projected disk midplane along the line of sight.

\begin{figure}
  \centering
  \includegraphics[scale=0.23]{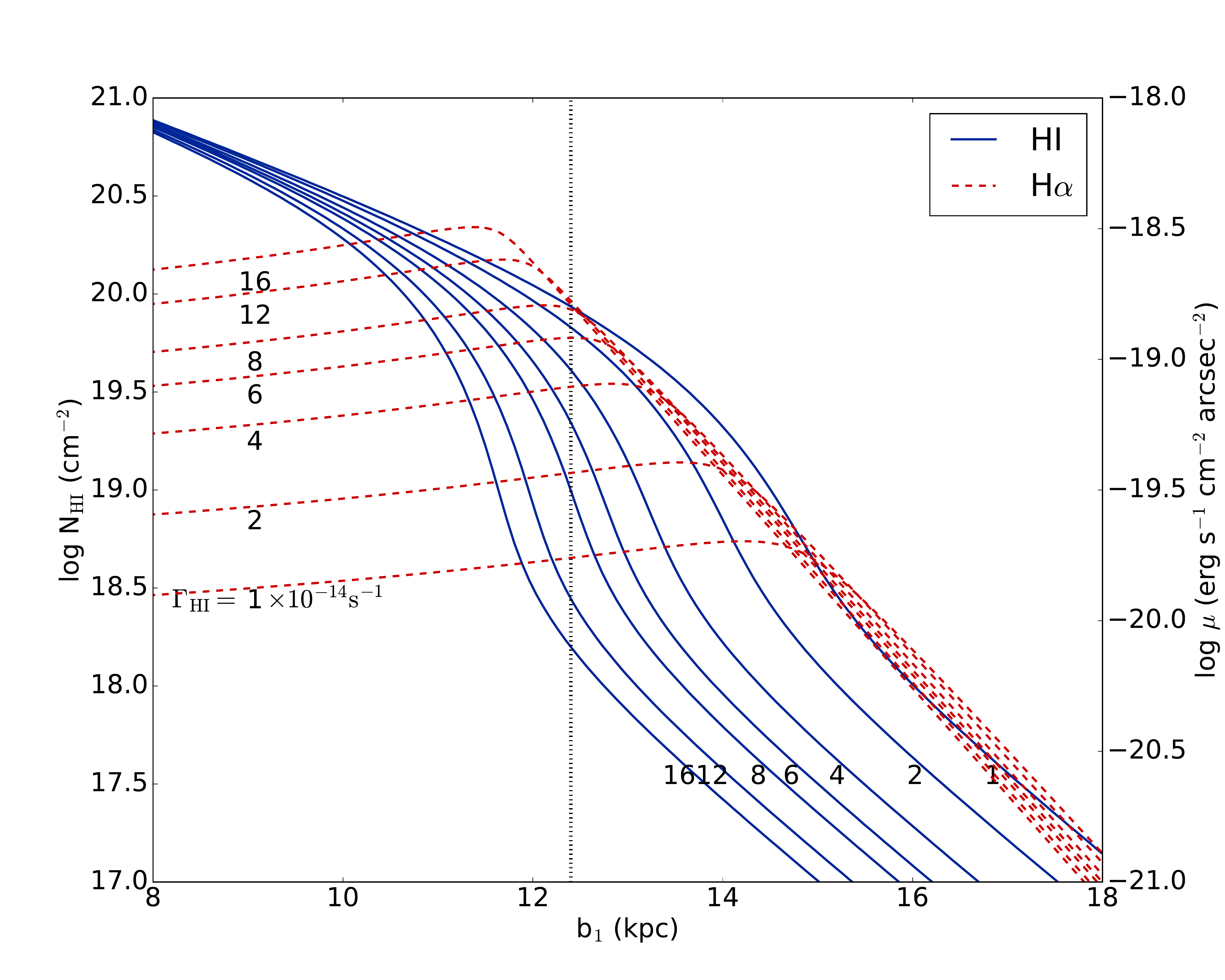}\\
  \caption{\HI\ profiles (blue solid lines, left hand-side axis) and \ha\ SB profiles
    (red dashed lines, right hand-side axis) extracted along the midplane ($b_2=0$) of
    a disk with constant parameters ($n_{\rm H,0}= 1.5~\rm cm^{-3}$, $h_{\rm R}=2300~\rm pc$,
    $h_{\rm z}=426~\rm pc$) observed at an inclination of $i = 84$ deg. Different curves are for
    different radiative transfer calculations with varying intensity of the UVB, as labelled by
    the respective values for $\Gamma_{\rm HI}$ in units of $10^{-14}~\rm s^{-1}$.
    The position of the $N_{\rm HI} = 10^{19}~\rm cm^{-2}$ contour for \gal\ is marked
    by the dotted black line. For a fixed density
    distribution, higher values of $\Gamma_{\rm HI}$ shift the ionisation front  to smaller $b_1$.
    The location of the maximum SB tracks the hydrogen ionisation front. The SB becomes nearly independent of $b_1$ at small radii, with an amplitude that is proportional to $\Gamma_{\rm HI}$.}\label{fig:cnstdis}
\end{figure}

\begin{figure}
  \centering
  \includegraphics[scale=0.23]{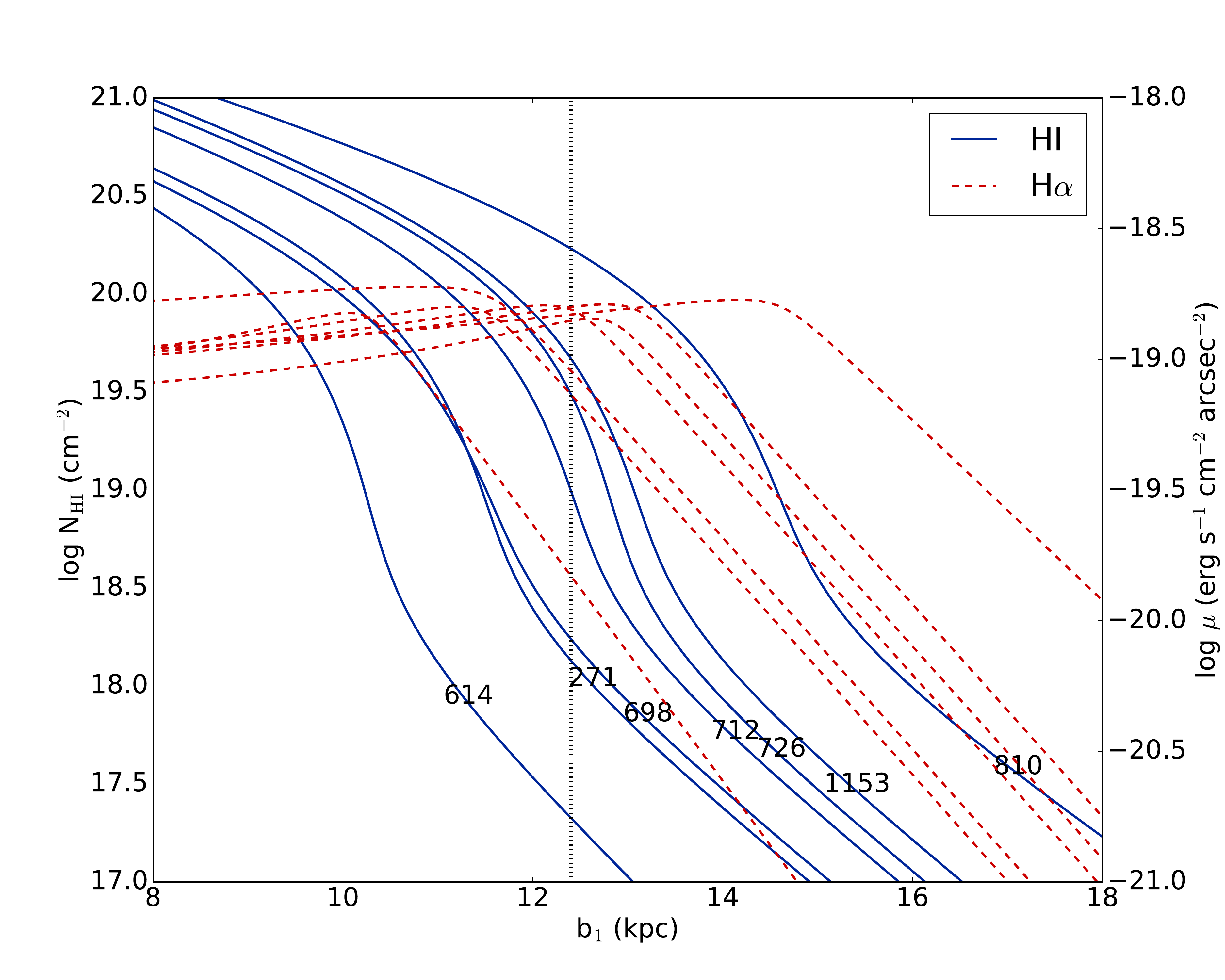}\\
  \caption{Same as Figure \ref{fig:cnstdis}, but for disks with different structural parameters (discussed in the
    text, curves are labelled with the identifier of the model as in Figure \ref{fig:cnstgamma2d}),
    which are illuminated by a constant UVB with $\Gamma_{\rm HI} = 8 \times 10^{-14}~\rm s^{-1}$. 
    Any inference on the properties of the UVB from \HI\ maps alone suffers from
    a degeneracy between the gas density profile and its ionisation state. However, the H$\alpha$ SB is only weakly dependent 
    on the density distribution for radii interior to the ionisation front.}\label{fig:cnstgamma}
\end{figure}

\begin{figure*}
  \centering
  \includegraphics[scale=0.6]{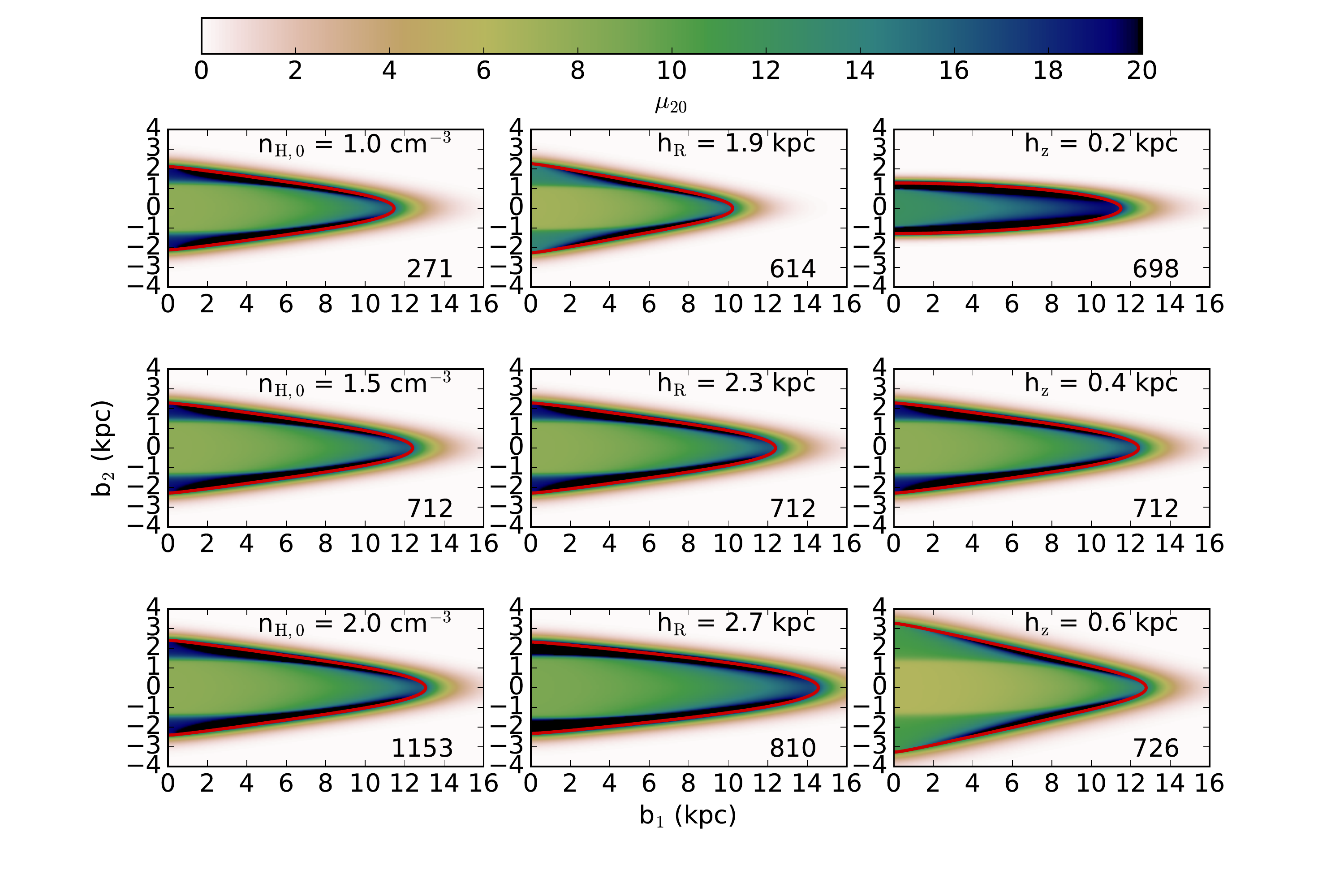}\\
  \caption{Gallery of \ha\ SB maps of a disk viewed at an inclination of $i = 84$ deg and illuminated by
    UVB with $\Gamma_{\rm HI} = 8 \times 10^{-14}~\rm s^{-1}$. Starting from a fiducial model
    with $n_{\rm H,0}= 1.5~\rm cm^{-3}$, $h_{\rm R}=2300~\rm pc$,
    $h_{\rm z}=426~\rm pc$ (model 712, in the middle row),  each column shows results
    of different disks obtained by varying structural parameters one at the time, as labelled in the top right
    corner of each panel. The red line marks the contour at $N_{\rm HI} = 10^{19}\rm ~cm^{-2}$. Profiles
    extracted from these models
    are shown in Figure \ref{fig:cnstgamma}, labelled by their ID number in the bottom right corner of each panel.
    By combining spatially resolved maps of \HI\ and \ha\ emission, one can discriminate
    among different models to accurately measure $\Gamma_{\rm HI}$.}\label{fig:cnstgamma2d}
\end{figure*}

\begin{figure*}
  \centering
  \includegraphics[scale=0.5]{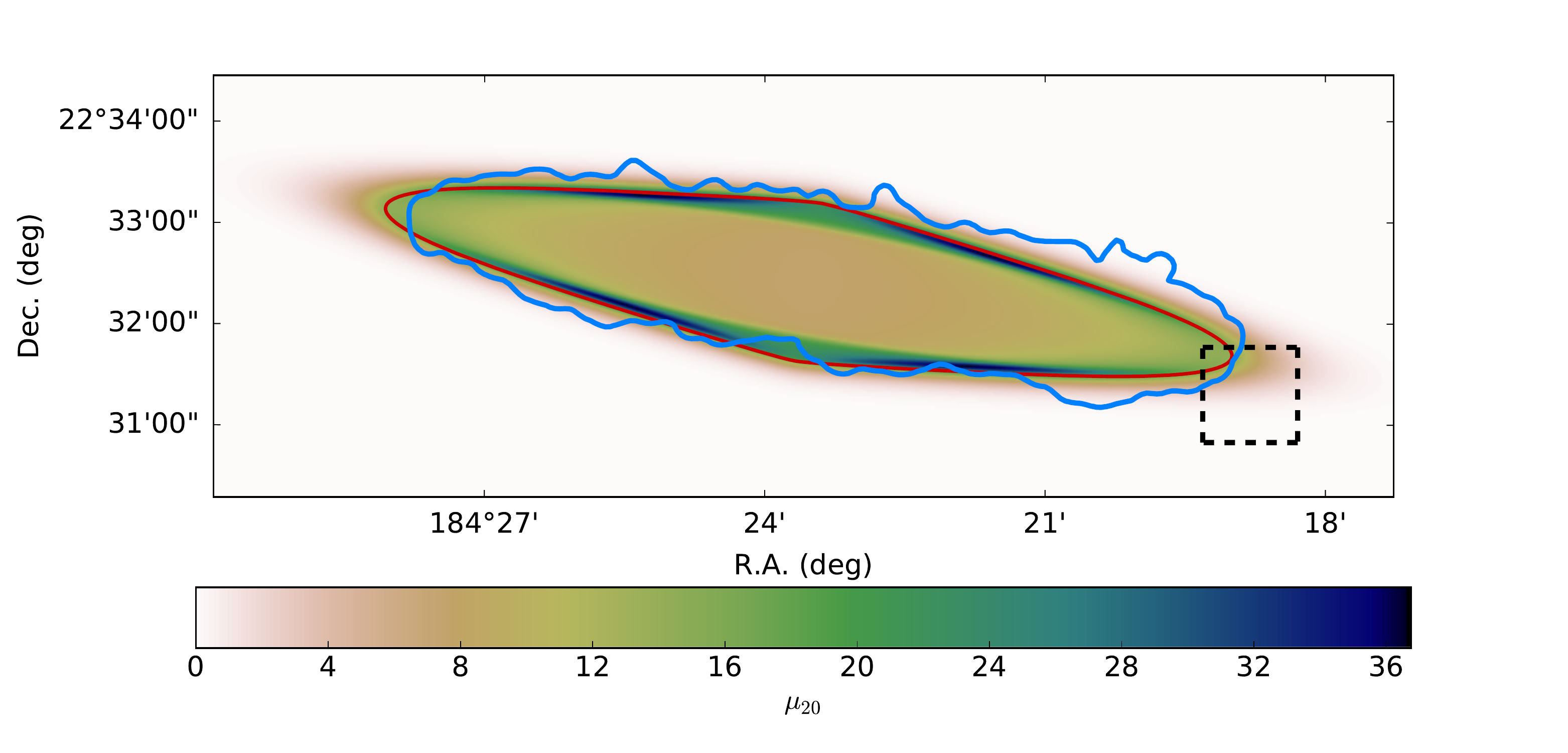}\\
  \caption{\ha\ SB map for the model in our grid that more closely reproduces
    both the observed \ha\ SB and the size of the \HI\ contours at $N_{\rm HI} = 10^{19}~\rm cm^{-2}$.
    The light blue line is the observed \HI\ contour at $N_{\rm HI} = 10^{19}~\rm cm^{-2}$, while the
    red line marks the location predicted by the model. The MUSE field of view is indicated by the black dashed line.
    Our observations constrain the photoionization rate to be in the range $\Gamma_{\rm HI} \sim (6-8) \times 10^{-14}~\rm s^{-1}$.
    However, due to the unknown contribution from local sources of ionisation, we caution that these values
    should be regarded as formal upper limits.}\label{fig:bestfit}
\end{figure*}

\subsubsection{Model predictions and general considerations}

Before turning our attention to the modelling of \gal, we present results from our radiative
transfer calculations to highlight how observations of the \HI\ column density combined
with observations of the \ha\ SB can constrain the \HI\ photoionization rate. 

In Figure \ref{fig:cnstdis}, we show the \HI\ column density and the corresponding \ha\ SB profiles computed along the midplane
of a projected disk ($b_2 = 0$), which is defined by $n_{\rm H,0}= 1.5~\rm cm^{-3}$, $h_{\rm R}=2300~\rm pc$,
and $h_{\rm z}=426~\rm pc$.
The disk is observed at an inclination of $i = 84$ deg, which is consistent with the inclination of \gal\ \citep{ada11}, and the different profiles are for different values of the \HI\ photoionization rate,
$\Gamma_{\rm HI} = (1,2,4,6,8,12,16) \times 10^{-14}~\rm s^{-1}$.
Trends that are common for this type of calculations can be found in this figure \citep[e.g.][]{dov94,ada11}.
Focusing on the \HI\ column density profiles, it is evident that the location of the ionisation front - where hydrogen
turns from highly ionised ($N_{\rm HI} \le 10^{17}~\rm cm^{-2}$) to fully neutral ($N_{\rm HI} \ge 10^{20}~\rm cm^{-2}$)
- moves to smaller radii with increasing \HI\ photoionization rate. The associated \ha\ SB profile
behaves similarly. The SB is maximal at the location of the ionisation front, drops off rapidly towards larger radii,
and slowly towards smaller radii.  Gas to the left of the ionisation front is
neutral along the midplane (for $\bar z =0$), and \ha\ emission arises from a skin of ionised gas above and below the
midplane that is observed in projection at $b_2 = 0$.  

Given that the shape of the \HI\ profiles varies with $\Gamma_{\rm HI}$,
observations of the location of this ionisation front in \HI\ maps
can be used to constrain the intensity of the UVB \citep[see also e.g.][]{dov94}. However, this measurement
is clearly degenerate with the structural parameters that define the gas density distribution.
This degeneracy is highlighted in Figure \ref{fig:cnstgamma}, which shows the \HI\ and \ha\ profiles
from a grid of models at constant $\Gamma_{\rm HI} = 8\times 10^{-14}~\rm s^{-1}$, but with different parameters
describing the structure of the disk. Starting with a fiducial model defined by $n_{\rm H,0}= 1.5~\rm cm^{-3}$,
$h_{\rm R}=2300~\rm pc$, $h_{\rm z}=426~\rm pc$ (ID 712),
we construct a grid varying each parameter one at the time, as shown in Figure \ref{fig:cnstgamma2d}.
All models are observed at an inclination angle of $84$ deg.
Comparing Figure \ref{fig:cnstdis} and Figure \ref{fig:cnstgamma}, it is clear that the location of the ionisation front is a sensitive function of the parameters describing
the density distribution of the disk, making accurate determinations of the UVB intensity from
\HI\ data alone very difficult. And while the \HI\ profiles in predominantly neutral regions
can be used to constrain the choice of structural parameters for a given galaxy,
co-variance among these parameters and$/$or local deviations of the density profile from a single exponential may lead
to incorrect extrapolations at larger radii, resulting in significant errors on the inferred photoionization
rate when using 21~cm data only.

However, Figure \ref{fig:cnstgamma} shows that the value of the \ha\ SB interior to the ionisation front is only weakly dependent on the disk's parameters, in spite of the large scatter in the location of the ionisation front itself or in the shape of the \HI\ profile. Thus, while \HI\ data alone provide only weak constraints on the photoionization rate, a joint analysis of the radial \HI\ column density and the \ha\ SB profile has the potential of pinning down $\Gamma_{\rm HI}$ to better than a factor of two for fiducial values of $\Gamma_{\rm HI} \sim (6-8) \times 10^{-14}~\rm s^{-1}$ (see below). Moreover, by exploiting MUSE's capability of obtaining spatially-resolved maps of the \ha\ SB, one can derive even tighter constraints on the photoionization
	rate through a joint analysis of the \HI\ and \ha\ maps in two dimensions. We do not attempt such a detailed analysis here, given
	the large uncertainty in the current line flux measurements. Thus, in the following, we simply offer a qualitative description of
	the advantages of resolving the spatial distribution of the \ha\ SB.

In Figure \ref{fig:cnstgamma2d}, we present the 2D maps for the same models shown in Figure \ref{fig:cnstgamma}. This gallery visually confirms  how models illuminated by a constant UVB with $\Gamma_{\rm HI} = 8\times 10^{-14}~\rm s^{-1}$ consistently reach a mean SB of $\sim 8 \times 10^{-20}$\sbline\ interior to the ionisation front. However, differences in the underlying density distribution result in characteristic shapes of the 2D SB maps. In particular, the location and shape of the brightest regions, which originate from projections effects of the ionisation front, are sensitive to the parameters describing the density distribution and, not shown here, to the viewing angle. It follows that precise determinations of the \HI\ photoionization rate are possible provided that one resolves these features in \ha\ SB maps, which can be analysed jointly with the \HI\ column density from 21 cm maps.

\subsection{The UVB photoionization rate at $z\sim 0$}\label{sec:uvblimit}

\subsubsection{Constraints on $\Gamma_{\rm HI}$}

Following the procedure outlined in the previous section, we combine information from the
\HI\ column density and the \ha\ SB maps to translate our measurement into a value of the
\HI\ photoionization rate. We start by constructing a grid of $\sim 5000$ radiative transfer models for exponential disks,
varying the central density in the interval $n_{\rm H,0}=1-6~\rm cm^{-3}$ in steps of $0.5~\rm cm^{-3}$,
the disk scale-length in the interval $h_{\rm R} = 1.3-2.9~\rm kpc$ in steps of $200~\rm pc$,
and the disk scale-height in the interval $h_{\rm z} = 100-700~\rm pc$ in steps of $100~\rm pc$.
These intervals are chosen to bracket the best-fitting parameters for \gal, as listed in table 1 of
\citet{ada11}. Similarly, the step size is chosen to be comparable to the statistical errors on these measurements. 
For each combination of disk parameters, we perform the radiative transfer calculation for seven different
values of the UVB intensity, $\Gamma_{\rm HI} = (1,2,4,6,8,12,16) \times 10^{-14}~\rm s^{-1}$.

Finally, \HI\ column densities and \ha\ SB maps are reconstructed projecting each model along three viewing angles
($i = 82,83,84$ deg), with values chosen to bracket the inclination of \gal\ in the plane of the sky as determined by \cite{ada11}. 
  We note that the inclination angle of \gal\ is uncertain, with  \citet{mat99} suggesting
  $i = 88$ deg \citep[see also][]{uso03}.
  This discrepancy reflects the difficulty of measuring inclinations for edge-on disks.
  In this work, we prefer to adopt a lower inclination angle, which appears to better reproduce the aspect ratio
  of the \HI\ disk for \gal\ at large radii, beyond the optical radius. We note
  that progressively higher inclination angles yield brighter and sharper ionization fronts,
  thus introducing an uncertainty in the inferred value for $\Gamma_{\rm HI}$ that is comparable to the error in
  the SB measurement.  As we discussed qualitatively in the previous section, future observations
  that can map the extent of the ionization front will be able to reduce this additional source of uncertainty.

Next, we select models that best reproduce the available observations by imposing the following two constraints
on the grid of projected \HI\ and \ha\ maps.
Firstly, we demand that the semi-major and semi-minor axes measured in models at $N_{\rm HI} = 10^{19}~\rm cm^{-2}$
match the observed values of $b_{1,{\rm HI}} = 12.4 \pm 0.1~\rm kpc$ and $b_{2,{\rm HI}} = 2.3 \pm 0.2~\rm kpc$
within the associated errors. Secondly, we require that the \ha\ SB computed in models at the same location
of our MUSE observations is consistent with the observed value within the associated errors. 
After imposing these two constraints, three models are found to provide a good fit to both the \HI\ and the
\ha\ observations. The best match, model ID 712, is shown in Figure \ref{fig:bestfit}.

The three models that best reproduce the available observations are characterised by disk parameters in the range
$n_{\rm H,0} = 1-2~\rm cm^{-3}$, $h_{\rm R} = 2.1-2.5~\rm kpc$, and $h_{\rm z} = 426~\rm pc$. The disk scale-length
and scale-height are consistent with the best-fitting parameters inferred by \citet{ada11}, obtained by modelling the
observed \HI\ profiles. Compared to their analysis, however, our models prefer smaller values for $n_{\rm H,0}$,
that, as shown in Figure \ref{fig:cnstgamma2d}, result in a smaller radius of the ionisation front.

This discrepancy arises from having imposed different constraints on the models. In their analysis, \citet{ada11} constrain their model to reproduce the \HI\ profile at radii between $9-11~\rm kpc$, and they extrapolate the best-fitting model to larger radii. Conversely, in our analysis we impose that the location of the $N_{\rm HI} = 10^{19}~\rm cm^{-2}$ contour in the model is also what is observed in the 21 cm map, without requiring that models track observations in the neutral regions at small radii.
	The fact that these different choices yield different best-fitting models is simply because the HI disk in the observed galaxy is not
	a perfect exponential. Given our choice and after simultaneously computing the properties of the \HI\ and \ha\ maps as a function of the \HI\ photoionization rate, our radiative transfer calculations predict that the maximal emission in \gal\ should occur in a region that
overlaps with the \HI\ region (see Figure \ref{fig:bestfit}), at radii further in than predicted by the model of
\citet{ada11}.  Indeed,  their figure 2 suggests that
an extrapolation of the best-fitting \HI\ profile is overestimating the radius of the \HI\ contour
at $N_{\rm HI} = 10^{19}~\rm cm^{-2}$ and, consequently, it is overestimating the extent of the
region with maximal SB. Conversely, our self-consistent calculation of the gas ionisation state and emissivity
shows that the lack of significant detection in the MAX region where the maximal emission was originally
expected is in fact fully consistent with a model in which gas at the edge of \gal\ is
photoionized by the UVB.

The models that more closely reproduce observations have \HI\ photoionization rates in the range of
$\Gamma_{\rm HI} \sim (6-8) \times 10^{-14}~\rm s^{-1}$.
We emphasise that this range does not represent a formal confidence interval on  $\Gamma_{\rm HI}$,
as it simply reflects the photoionization rates of models that are present in our grids and that
provide a good description of our observations. A more formal estimate of the photoionization rate
can be obtained, for instance, coupling the results of our radiative transfer code with Markov chain Monte Carlo
methods. However, given the large uncertainty currently affecting the measured SB, we defer this approach
to future work.

\begin{figure*}
  \centering
  \includegraphics[scale=0.5]{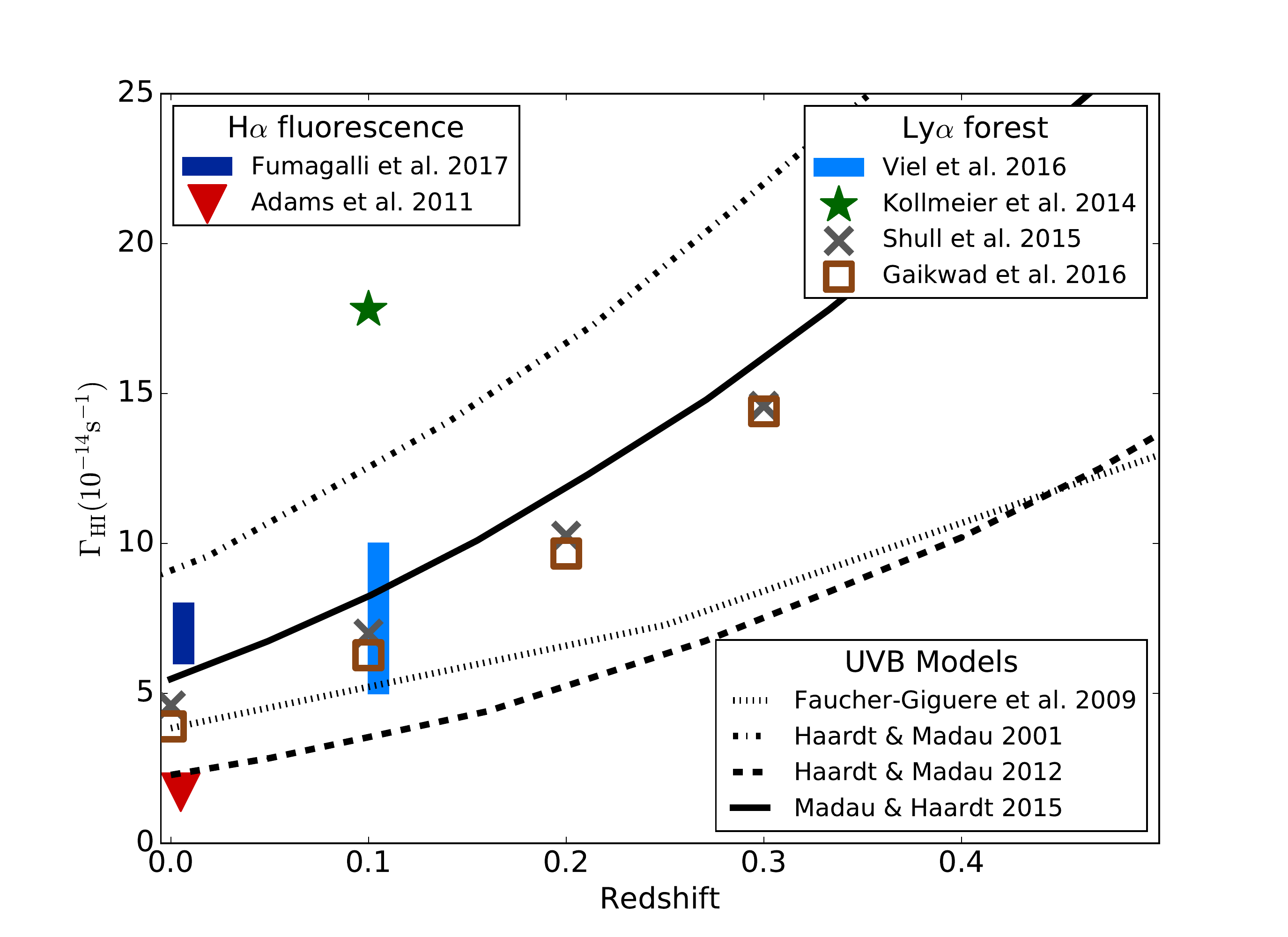}\\
  \caption{Summary of current models and constraints of UVB photoionization rate at $z<0.1$.
    Predictions from UVB models from \citet{haa01}, \citet{fau09}, \citet{haa12}, and \citet{mad15}
    are displayed with lines.
    Values inferred from statistics of the Ly$\alpha$ forest from \citet{kol14}, \citet{shu15},
    \citet{vie16}, and \citet{gai16} are shown with symbols.
    The upper limit inferred by \citet{ada11} is also shown in comparison with our measurement.}\label{fig:gammaz}
\end{figure*}

\subsubsection{Caveats on ionisation mechanisms other than the UVB}

Before comparing our findings with previous work, we note that the inferred value of $\Gamma_{\rm HI}$ should be
regarded as a formal upper limit on the UVB photoionization rate. 
Indeed, in addition to recombination from gas photoionized by the UVB, \ha\ emission in proximity
to a galaxy may arise from photoionization due to local sources 
or from ionising photons that escape from the inner star-forming disk \citep[e.g.][]{vog06,oey07}.
Furthermore, processes other than photoionization may operate at the disk-halo transition, as seen for instance
within the diffuse ionised medium in nearby galaxies \citep[e.g.][]{hoo03,cal04}.
 With current data, we cannot easily constrain the nature of the ionization mechanism,
   as detailed modelling would require, for instance, the detection of metal lines in deeper exposures \citep[e.g.][]{bla97}.
We note, however, that a significant contribution from local sources is unlikely, as MUSE observations would in fact resolve
HII regions with sizes of $\gtrsim 30$ pc. Even accounting for smaller unresolved HII regions
at the position of our observations, we do not expect star formation 
on scales of $\gtrsim 10''$ (or $\gtrsim 500~\rm pc$) in the outer \HI\ disk of \gal.

To assess instead the contribution of ionising photons from the central star-forming regions to the total
ionisation budget, we use the {\sc starburst99} code \citep{lei99} to generate a spectral energy distribution
$L_{\nu}$ for \gal. In this calculation, we assume a star formation rate of $\sim 0.6~\rm M_\odot~yr^{-1}$
based on the observed UV flux \citep{kar13}. The maximum contribution of local sources to 
the photoionization rate at the position of our observations is 
\begin{equation}
\Gamma_{\rm HI,loc} = \int^{\infty}_{\nu_0} d\nu \frac{L_\nu}{4\pi R^2h\nu}\sigma_{\rm HI}(\nu) \sim 2.6\times 10^{-11}~\rm s^{-1}\:,
\end{equation}
where the numerical value for $\sigma_{\rm HI}$ is from \citet{ver96}, and
$R = 12.5~\rm kpc$ is the distance from the observed position to the centre of the galaxy.

This calculation \citep[see also][]{sch06} implies that massive stars in the disk of \gal\ can easily
account for, and exceed, the photoionization rate inferred by our observations. 
However, by targeting the tip of the \HI\ disk, we  maximise the optical depth seen by ionising photons that
leak from the central stellar disk. As shown by the locations of the \HI\ contours in Figure \ref{fig:galfov}, photons leaking
along the midplane would see an optical depth at $\lambda \sim 912~$\AA\ of $\tau_{912} \gg 1000$
and, although the small-scale structure of the ISM is likely to be very different from that of a simple slab, 
the presence of column density in excess of $N_{\rm HI} = 10^{21}~\rm cm^{-2}$ makes it quite unlikely that ionising photons escape along the disk's midplane.
The assumed geometry for the calculation of the optical depth at the midplane is also justified, to first order, by the fact that \gal\ does not exhibit a prominent central bulge and that its disk does not present notable irregularities in the stellar or \HI\ distribution \citep{mat99,uso03}.

We conclude that the \ha\ emission at the location of our observations is primarily driven by photoionization arising from the UVB. We think it is unlikely that other sources contribute significantly but cannot rule out that ionising photons from the galaxy itself or from other sources contribute as well. Therefore we can only place a formal upper limit on the \HI\ photoionization rate of the UVB with current data, but we regard our measurement at the edge of the \HI\ disk as a bona-fide estimate of the intensity of the ionising UVB. 

\subsubsection{Comparison with other work}

In Figure \ref{fig:gammaz}, we compare our inferred values for $\Gamma_{\rm HI}$ with
predictions from models of the UVB and with other estimates from the recent literature. 
Our measurement, taken at face value, is in disagreement with the $5\sigma$ upper limit on $\Gamma_{\rm HI}$ reported
by \citet{ada11} in the same galaxy \gal, despite their SB upper limit being consistent with our detection.
This discrepancy highlights how
detailed radiative transfer calculations are required when converting the \ha\ SB into a photoionization rate.
As the quality of SB measurements are likely to improve in the near future (see below), more detailed modelling is
therefore warranted to characterise the systematic uncertainty that affects the conversion between observables ($\mu$) and physical
quantities ($\Gamma_{\rm HI}$). Considering instead the photoionization rates inferred from the statistics of the low-redshift
($z\sim 0.1$) Ly$\alpha$ forest, our values for $\Gamma_{\rm HI}$ are in line with the recent analyses by
\citet{shu15}, \citet{vie16}, and \citet{gai16}. These authors consistently find values in the
interval $\Gamma_{\rm HI} \sim (5-10)\times 10^{-14}~\rm s^{-1}$ \citep[but see][]{kol14},
albeit relying for most part on the analysis of the same data.
Extrapolated to $z\sim 0$, these studies predict $\Gamma_{\rm HI} \sim (4-5)\times 10^{-14}~\rm s^{-1}$, below but
broadly consistent with our determination of $\Gamma_{\rm HI} \sim (6-8)\times 10^{-14}~\rm s^{-1}$.

Turning our attention to models of the UVB, we note that cosmological radiative transfer calculations predict
photoionization rates that bracket current measurements.
Indeed, both our measurement and the study of the
$z\sim 0.1$ Ly$\alpha$ forest imply \HI\ photoionization rates at intermediate values relative to those predicted by the
\citet{haa12} and \citet{fau09} models (at the lower end) and by the \citet{haa01} model (at the upper end).
Figure \ref{fig:gammaz} also suggests that, as already discussed in the literature \citep[e.g.][]{kol14,shu15,kha15},
the recent \citet{haa12} UVB model may underestimate $\Gamma_{\rm HI}$ by a factor of $\sim 2-3$.
The new \citet{mad15} model, which has been recalculated
with updated quasar emissivity \citep[see also][]{kha15,cri16}, lies instead in the range allowed by
observations.
However, the large scatter
among measurements both at $z\sim 0$ \citep[e.g compare our value and the limit by][]{ada11} and at $z\sim 0.1$
\citep[e.g. compare][]{kol14,shu15} imply that current measurements still suffer
from up to a factor $\sim 2$ uncertainty, and better accuracy is needed to further inform
and refine models. 

\section{Summary and Future Prospects}\label{sec:end}

We have presented new MUSE observations targeting the edge of the \HI\ disk in the nearby edge-on galaxy
\gal. An emission line is detected in a deep 5.7-hour exposure  at $\lambda \sim 6574~$\AA,
which is the wavelength where H$\alpha$ is expected given the \HI\ radial velocity of \gal.
The emission line is also spatially resolved in narrow band images
reconstructed from the MUSE data cube. The detected signal is located in close proximity of the edge of the MUSE FOV,
and it lies in the wing of a sky line at $\lambda \sim 6577~$\AA. Combined, these effects introduce a
substantial uncertainty that dominates the error budget of our measurement.

Despite these additional sources of uncertainty, we have shown that an astrophysical signal is consistently recovered within
data cubes reduced with different pipelines, and within data cubes containing two independent sets of exposures.
Further, through the study of mock data cubes, we have shown that the detected emission line has properties consistent with that
expected from an astrophysical signal associated with \gal. Altogether, we conclude that we have detected \ha\ recombination
from the edge of the \HI\ disk of \gal\ with a line SB of $(1.2 \pm 0.1 \pm 0.5) \times 10^{-19}~$\sbline.
Here, the first error indicates the statistical uncertainty and the second error characterises the presence
of the additional sources of uncertainties discussed above.

We present new radiative transfer calculations that self-consistently solve for the ionisation and temperature
structure of an exponential disk. The joint analysis of spatially-resolved
\HI\ column density and \ha\ SB maps enables us to translate the observed SB
into a value for the \HI\ photoionization rate of the UVB.
Following this procedure, our current measurement
implies $\Gamma_{\rm HI} \sim (6-8)\times 10^{-14}~\rm s^{-1}$, which is in line with the values inferred from the statistics of
the low-redshift Ly$\alpha$ forest. 
While it is quite likely that H$\alpha$ emission at the location of our observations is primarily driven by photoionization arising
from the UVB, we caution that an unknown contamination from other sources of ionization may be present. Thus, we can only place a formal
upper limit on the \HI\ photoionization rate with current observations, but we consider this measurement at the edge of the
\HI\ disk as an estimate of the actual intensity of the ionizing UVB.

Despite the substantial systematic uncertainty that affects our measurement, our work has demonstrated the potential
that future MUSE observations have in constraining the intensity of the UVB in the local Universe.
Through a grid of radiative transfer calculations, we have shown how detailed \ha\ maps of the ionisation
front could be used jointly with \HI\ maps to precisely constrain $\Gamma_{\rm HI}$.
To achieve better measurements of the UVB intensity, future work should however address the following
two limitations of the current analysis.

Observationally, better precision on the SB measurement is mandatory for
improving constraints on the UVB photoionization rate. With the use of mock data, we have shown that
a precision of $\sim 5-10\%$ can be easily achieved with MUSE in regions close to the centre of the FOV and
far from sky lines. Thus, thanks to improved predictions of the spatial location of the \ha\ emission in \gal\
(Figure \ref{fig:bestfit}), MUSE follow-up observations are expected to sample more accurately the location of the ionisation
front. Also, by observing the East side of the galaxy with a radial velocity of
$\sim 300-400~\rm km~s^{-1}$, the \ha\ emission line  shifts to $\sim 6570~$\AA, in a region away
from sky lines. Thus, measurements with errors below $\sim 5-10\%$ should be possible in the near future.

As the precision of observations improves, models should be refined to
reliably convert the observed SB into a measurement for the \HI\ photoionization
rate. The primary effect that should be accounted for in future analyses is the
impact of local sources of ionisation. Models that include only the effects of the UVB predict
a characteristic shape of the \ha\ emission (Figure \ref{fig:cnstgamma2d}) that can be used to test whether
the edge of the disk is illuminated by an external radiation field or whether local sources
contribute significantly.  
Ancillary multiwavelength observations for the star-forming disk of \gal\ should be
used to constrain the spectral energy distribution of local sources that can be added
as further contribution to the ionisation budget in radiative transfer models. 
Additional improvements include the treatment of metals and dust
in a full three-dimensional radiative transfer calculation.

To further obviate to the problem of local sources, MUSE observations can target ``dark'' clouds,
where prominent star formation is absent.
This experiment has been already attempted, for instance,
targeting the intergalactic cloud \HI~1225$+$01 \citep[e.g.][]{vog95,wey01}. 
Deep MUSE observations will be able to further improve on current limits on the \ha\ SB, reaching
levels of $\sim 10^{-20}~$\sbline. Future experiments could also search for the population of ``RELHICs'',
which are dark and gas-rich halos that are predicted in ${\Lambda}$CDM simulations \citep[e.g.][]{ste02,dav06,ben16}.
Given the simple physics that regulates the properties of these dark galaxies, accurate measurements of the UVB
intensity should be possible with deep MUSE follow-up observations.

From our analysis and from these considerations, we conclude that
new measurements of the UVB intensity at $z\sim 0$ via \ha\ fluorescence appear within reach
in the era of large format integral field spectrographs at 8m class telescopes.

\section*{Acknowledgements}
We are grateful to Joshua Adams for sharing \HI\ and \ha\ contours from \citet{ada11} in electronic 
form. We thank Xavier Prochaska, Joshua Adams, Juan Uson, and an anonymous referee for
useful comments that have improved our work.
M Fumagalli, TT, and SM acknowledge support by the Science and Technology Facilities Council [grant number  ST/L00075X/1]. 
Support for this work was provided to PM by NASA through grant HST-AR-13904.001-A. PM also acknowledges a NASA contract 
supporting the WFIRST-EXPO Science Investigation Team (15-WFIRST15-0004), administered by GSFC, and thanks the
Pr\'{e}fecture of the Ile-de-France Region for the award of a Blaise Pascal International Research Chair, managedby the
Fondation de l'Ecole Normale Sup\'{e}rieure. SC gratefully acknowledges support from Swiss National Science Foundation
grant PP00P2\_163824. M Fossati acknowledges the support of the Deutsche Forschungsgemeinschaft via
Projects WI 3871/1-1, and WI 3871/1-2.  This work is based on observations collected at the European Organisation
for Astronomical Research  in the Southern Hemisphere under ESO programme ID 095.A-0090. 
This research made use of Astropy, a community-developed core Python package for Astronomy \citep{astropy}.
For access to the data and codes used in this work, please contact the authors or visit 
\url{http://www.michelefumagalli.com/codes.html}. 


\label{lastpage}

\end{document}